\documentclass[12pt]{article}

\pdfoutput=1 
\usepackage[T1]{fontenc} 
\usepackage{amsmath,amssymb,amsfonts}
\usepackage[utf8]{inputenc}
\usepackage{graphicx}
\usepackage{booktabs,colortbl}
\usepackage{cancel,xcolor,multirow}
\usepackage[normal]{caption}
\usepackage{subcaption}
\usepackage{slashed}
\usepackage{pifont}
\usepackage{cite}
\usepackage{hyperref}
\hypersetup{%
  colorlinks = true,
  linkcolor  = black
}
\newcommand{\tev}{~\text{TeV}}
\newcommand{\gev}{~\text{GeV}}

\newcommand{\Lc}{\mathcal{L}}

\newcommand{\drho}{{\rho_D}}
\newcommand{\dpi}{{\pi_D}}

\newcommand{\eps}{\epsilon}
\newcommand{\ra}{\rightarrow}
\newcommand{\lsim}{\lesssim}

\DeclareUrlCommand\ULurl{%
  \renewcommand\UrlLeft{\uline\bgroup}%
  \renewcommand\UrlRight{\egroup}}
  
\hypersetup{colorlinks=true,urlcolor=blue, citecolor=blue}
\newcommand{\GeV}{~\mathrm{GeV}}

\begin{document}

\pagestyle{plain}


\begin{center}
{\Large Dark Mesons at the LHC}

\vspace{1cm}

{Graham D. Kribs$^1$, Adam Martin$^2$, Bryan Ostdiek$^1$, and Tom Tong$^1$}
\end{center}

\vspace*{0.5cm}

\noindent
\hspace*{1.3cm} $^1$Department of Physics, University of Oregon, Eugene, OR 97403 \\
\hspace*{1.3cm} $^2$Department of Physics, University of Notre Dame, South Bend, IN 46556 \\

\vspace*{0.5cm}

 
\begin{abstract}

A new, strongly-coupled ``dark'' sector could be accessible 
to LHC searches now.  These dark sectors consist of composites 
formed from constituents that are charged under the electroweak group 
and interact with the Higgs, but are neutral under Standard Model color. 
In these scenarios, the most promising target is the dark meson sector,
consisting of dark vector-mesons as well as dark pions.
In this paper we study dark meson production and decay at the LHC
in theories that preserve a global $SU(2)$ dark flavor 
symmetry. 
Dark pions -- like the pions of QCD -- 
can be pair-produced through resonant dark vector meson production,
$p p \rightarrow \rho_D \ra \pi_D \pi_D$, and decay in 
one of two distinct ways:  
``gaugephobic'', when $\pi_D \ra f \bar{f}'$ generally dominates; 
or ``gaugephilic'',  when $\pi_D \ra W + h, \, Z + h$ dominates
once kinematically open. 
Unlike QCD, the decay $\pi^0_D \ra \gamma\gamma$ is virtually absent
due to the dark flavor symmetry.
We recast a vast set of existing LHC
searches to determine the current constraints on
(and future opportunities for) dark meson production
and decay.  When $m_{\rho_D}$ is slightly heavier than $2 m_{\pi_D}$ and 
$\rho_D^{\pm,0}$ kinetically mixes with the weak gauge bosons,
the $8$~TeV same-sign lepton search strategy sets the best bound, 
$m_{\pi_D} > 500$~GeV\@.  
Yet, when only the $\rho^0_D$ kinetically mixes with hypercharge,
we find the strongest LHC bound is $m_{\pi_D} > 130$~GeV, that 
is only slightly better than
what LEP~II achieved two decades ago.  We find the relative 
insensitivity of LHC searches, especially at $13$~TeV, can be 
blamed mainly on
their penchant for high mass objects or large missing energy.
Dedicated searches would undoubtedly yield substantially improved sensitivity.
We provide a GitHub page to speed the implementation of  
these searches in future LHC analyses.
Our findings for dark meson production and decay provide a strong
motivation for {\em model-independent} searches of the form
$pp \rightarrow A \rightarrow B + C 
    \rightarrow {\rm SM} \, {\rm SM} + {\rm SM} \, {\rm SM}$
where the theoretical prejudice is for ${\rm SM}$ to be a 
3rd generation quark or lepton, $W,Z$, or $h$.  

\end{abstract}

\newpage

\setlength{\parskip}{0em}
\tableofcontents
\setlength{\parskip}{1em}


\section{Introduction}
\label{sec:Introdcution}

We consider extensions of the Standard Model that incorporate
a new, strongly-coupled, confining gauge theory with 
fermion representations that transform under the
electroweak group. 
There are a wide variety of uses of a new, strongly-coupled, 
confining group.  One use is to at least partially break 
electroweak symmetry dynamically, such as bosonic technicolor~\cite{Simmons:1988fu,Samuel:1990dq,Dine:1990jd,Kagan:1990az,Kagan:1991gh,Carone:1992rh,Carone:1993xc,Dobrescu:1997kt,Antola:2009wq} 
and the closely related ideas on strongly-coupled
induced electroweak symmetry breaking~\cite{Azatov:2011ht,Azatov:2011ps, Gherghetta:2011na,Galloway:2013dma,Chang:2014ida,Beauchesne:2015lva,Harnik:2016koz,Alanne:2016rpe,Galloway:2016fuo,Agugliaro:2016clv,Barducci:2018yer}. 
Composite Higgs theories also posit a new strongly-coupled sector 
in which at least an entire Higgs doublet emerges in the 
low energy effective theory (the literature is far too vast to 
survey, for a review see e.g., \cite{Bellazzini:2014yua}). 
There is also a interesting connection to the
relaxation of the electroweak scale \cite{Graham:2015cka}
using a new strongly-coupled sector, e.g., 
\cite{Graham:2015cka,Antipin:2015jia,Agugliaro:2016clv,Batell:2017kho,Barducci:2018yer}.

Dark matter can emerge as a composite meson or baryon of a
strongly-coupled theory, often with an automatic accidental symmetry 
that protects against its decay. 
Since the early days of technicolor there was a possibility of 
dark matter emerging as technibaryons
\cite{Nussinov:1985xr,Chivukula:1989qb,Barr:1990ca,Barr:1991qn,Kaplan:1991ah,Chivukula:1992pn,Bagnasco:1993st}.  
There is now a growing literature
that has studied strongly-coupled dark matter as dark pions
\cite{Khlopov:2008ty,Ryttov:2008xe,Hambye:2009fg,Bai:2010qg,Lewis:2011zb,Frigerio:2012uc,Buckley:2012ky,Bhattacharya:2013kma,Marzocca:2014msa,Hietanen:2014xca,Pasechnik:2014ida,Antipin:2014qva,Hochberg:2014kqa,Carmona:2015haa,Lee:2015gsa,Hochberg:2015vrg,Bruggisser:2016ixa,Ma:2017vzm,Davoudiasl:2017zws,Berlin:2018tvf,Choi:2018iit,Hochberg:2018rjs}, 
dark quarkonia-like states
\cite{Alves:2009nf,Kribs:2009fy,Lisanti:2009am,Alves:2010dd,Geller:2018biy},
as well as dark baryons and related candidates 
\cite{Gudnason:2006yj,Dietrich:2006cm,Foadi:2008qv,Khlopov:2008ty,Mardon:2009gw,Kribs:2009fy,Sannino:2009za,Barbieri:2010mn,Belyaev:2010kp,Lewis:2011zb,Appelquist:2013ms,Hietanen:2013fya,Cline:2013zca,Appelquist:2014jch,Krnjaic:2014xza,Hietanen:2014xca,Detmold:2014qqa,Detmold:2014kba,Brod:2014loa,Asano:2014wra,Appelquist:2015yfa,Appelquist:2015zfa,Drach:2015epq,Fichet:2016clq,Co:2016akw,Dienes:2016vei,Ishida:2016fbp,Francis:2016bzf,Lonsdale:2017mzg,Berryman:2017twh,Mitridate:2017oky,Francis:2018xjd} (for a review, see \cite{Kribs:2016cew}).

Another use is to simply characterize generic strongly-coupled-like 
signals as targets for LHC and future colliders. 
Vector-like confinement \cite{Kilic:2009mi} pioneered this study
in the context of vector-like fermions that transform under
part of the SM group as well as under a new, strongly-coupled group
with scales near or above the electroweak scale.
Further explorations into the phenomenology and especially the
meson sector included
\cite{Kribs:2009fy,Kilic:2010et,Harnik:2011mv,Fok:2011yc,Buckley:2012ky,Bai:2013xga,Brod:2014loa,Chacko:2015fbc,Agashe:2016rle,Matsuzaki:2017bpp,Barducci:2018yer,Draper:2018tmh,Buttazzo:2018qqp}.  
In theories with somewhat lower confinement scales,
the dark sector may lead to dark showers and related
phenomena~\cite{Schwaller:2015gea,Cohen:2015toa,Freytsis:2016dgf,Zhang:2016sll,Cohen:2017pzm,Beauchesne:2017yhh,Renner:2018fhh},
displaced signals \cite{Mahbubani:2017gjh,Buchmueller:2017uqu}
and potentially intriguing spectroscopy
\cite{Hochberg:2015vrg,Daci:2015hca,Hochberg:2017khi}.
Spectacular ``quirky'' signals can arise in theories 
with a very low confinement scale \cite{Han:2007ae,Kang:2008ea}. 
The latter theories may also lead to a high multiplicity of
soft particles that are tricky to observe
\cite{Harnik:2008ax,Knapen:2016hky,Pierce:2017taw}.

In a companion paper \cite{Kribs:2018oad}, we develop dark sectors whose
(ultraviolet) strongly-coupled sector preserves a $SU(2)$ dark 
flavor symmetry.
These theories are mapped into a low energy effective theory that 
provides the leading interactions of the dark mesons with the 
Standard Model.  
A dark sector that is of particular interest to us 
is Stealth Dark Matter \cite{Appelquist:2015yfa}.
In this theory, there is a new, strongly-coupled dark sector
that consists of vector-like fermions that transform under both the 
new dark group as well as the electroweak part of the SM, 
and crucially, also permit Higgs interactions
(from Yukawa couplings or higher dimensional operators). 
Others have also pursued dark sectors with vector-like fermions
that permit Higgs interactions for a variety of purposes \cite{Antipin:2014qva,Antipin:2015jia,Beylin:2016kga,
Mitridate:2017oky,Barducci:2018yer}.

The dark meson sector of the Stealth Dark Matter theory
has several intriguing properties due to the 
accidental symmetries of the model.
Like vector-like confinement \cite{Kilic:2009mi}
the dark sector is free of constraints from 
precision electroweak observables and Higgs coupling
measurements so long as the vector-like mass is dominant.  
Unlike vector-like confinement, however, the Higgs interactions
break the global (species) symmetries of the dark sector, permitting
dark pions to decay into SM states.
Provided the vector-like masses are smaller than $\sim 4\pi f$, where $f$ 
is the scale of the new strong interaction, we can organize the states 
using chiral perturbation theory.  In this paper we focus on 
the most phenomenologically relevant states: 
the (lightest) triplet of pseudoscalar pions $\pi_D^a$ and the 
heavier triplet of vector mesons $\rho_D^a$
\cite{Ecker:1988te,Ecker:1989yg}. 
The scales of the theory, as we will see, are comparable to
or somewhat larger than the electroweak scale.

The presence of a $SU(2)$ dark flavor symmetry arises from
global symmetries of the ultraviolet strongly-coupled sector.  
For example, a strongly-coupled sector that contains 
two flavors of dark fermions with identical (current) masses
has a global $SU(2) \times SU(2)$ symmetry that is broken by the
condensate to a $SU(2)$ dark flavor symmetry \cite{Kribs:2018oad}.
This is just like QCD with its two light flavors of quarks with
nearly equal (current) masses.   
In Ref.~\cite{Kribs:2018oad}, we demonstrate strongly-coupled 
theories where the $SU(2)$ dark flavor symmetry can be 
identified as an exact custodial symmetry of the dark sector. 
That is, the Higgs multiplet interacts with the dark flavors
such that the $SO(3) \sim SU(2)_c$ is not further broken by the
dark sector.  Consequently, the dark sector's meson degrees of 
freedom can be categorized in custodial symmetric 
representations.
Again considering the example of theories with two flavors of 
dark fermions, the meson sector contains dark pions and 
one set of dark vector mesons in a triplet representation of the 
$SU(2)$ dark flavor symmetry.
Unlike QCD, however, the vector-like nature of the dark sector 
permits two possibilities for gauging the global flavor symmetry:   
the entire $SU(2)$ could be gauged (the $SU(2)_L$ weak interaction)
or just the $U(1)$ (as in $U(1)_B$ hypercharge).\footnote{It is 
also possible that there is some mixture
between $SU(2)_L$ and $SU(2)_R$, but this requires more than
just a single triplet of dark pions and dark vector mesons.
More details can be found in Ref.~\cite{Kribs:2018oad}.}
This leads to
two distinct low energy effective theories of dark mesons:
\begin{equation}
\begin{array}{rcl}
SU(2)_L \;\; {\rm model}: & \; & 
    SU(2)_{\rm global \; flavor} \leftrightarrow SU(2)_L \\ 
SU(2)_R \;\; {\rm model}: & \; & 
   SU(2)_{\rm global \; flavor} \leftrightarrow SU(2)_R 
\end{array}
\label{eq:models}
\end{equation}
In the latter case, obviously only the $U(1)$ subgroup is gauged, 
but since we assume the dark sector respects the full 
global $SU(2)$, we'll refer to this as the $SU(2)_R$ model.

In the meson sector the
dark pion states can be pair-produced, either via Drell-Yan or 
resonantly via mixing of the $\rho$ with SM electroweak gauge bosons. 
The dark pion decays can be categorized into two distinct possibilities:
``gaugephobic'', when $\pi_D \rightarrow f \bar{f}'$ dominates; 
or ``gaugephilic'',  when $\pi \rightarrow W + h, \, Z + h$ dominates
once kinematically open.  The decay $\pi^0_D \ra \gamma\gamma$ 
is highly suppressed due to the dark flavor symmetry \cite{Kribs:2018oad}.
For a wide range of parameters, the interaction between single
dark pions and the SM is small enough to make single pion production 
phenomenologically irrelevant, and yet, the interaction can be
easily large enough that the dark pions decay \emph{promptly} 
back to SM states.  We also briefly comment on the possibility that
dark pions are sufficiently long-lived so as to modify 
their phenomenological signature.

Dark mesons are therefore an example of new physics that must be pair 
produced with $\sim$ weak strength and decay back to multiple SM particles 
(only). The combination of a relatively low production cross section and 
complex final states with no BSM sources of missing energy leads to weak 
LHC constraints. We perform a detailed breakdown of which LHC searches 
could potentially set bounds on dark mesons. For the searches with 
potential, we recast the searches and estimate the bounds for some 
benchmark dark meson scenarios. For the searches that fail, we identify 
why. This latter step is useful as we find many 13 TeV analysis are 
insensitive to dark mesons because their cut thresholds are too high.

The layout of the rest of this paper is as follows. In Sec.~\ref{sec:phenodarkmesons} we introduce our phenomenological dark meson model and its relevant parameters. This model description is broken up into three parts: the strong sector, kinetic mixing, and $\pi_D$ decay. Using this setup, we explore the constraints on dark meson parameter space. Sec.~\ref{sec:collider} is devoted to constraints from single $\rho_D$ production, while we explore constraints from $\pi_D$ pair production in Sec.~\ref{sec:Resonant}. We step through the details of the searches that provide constraints and provide insight into why other searches fail to. Finally, we present our conclusions in Sec.~\ref{sec:conclusions}.

\section{Phenomenological Description of Dark Mesons}
\label{sec:phenodarkmesons}

The dark meson interactions will be described below using a 
phenomenological lagrangian.  The core philosophy was
formulated in ``vector-like confinement'' 
\cite{Kilic:2009mi,Kilic:2010et}, and our discussion
of resonant production of dark pions through a dark rho 
parallels theirs.  The key distinction
between our formulation and vector-like confinement is 
the presence of Higgs interactions among the dark fermions
which breaks enough of the dark flavor symmetries to allow 
dark pions to decay.  In the language
of vector-like confinement, all species symmetries are broken
by Higgs interactions in the dark sector (either
Yukawa couplings or higher-dimensional interactions).

\subsection{Dark Mesons in $SU(2)$ Triplet Representations}
 
The lagrangian can be written as
\begin{eqnarray}
\mathcal L &=& \mathcal L_{\rm strong} 
               + \mathcal L_{\rm kinetic \; mixing} 
               + \mathcal L_{\rm decay} \, . 
\label{eq:totlag}           
\end{eqnarray}
The first contribution contains the meson sector of the theory  
as it arises from the strongly-coupled dark sector:
\begin{eqnarray}
\mathcal L_{\rm strong} &=&{} 
  - \frac{1}{4} \drho_{\mu\nu}^{a}\drho^{a{\mu\nu}} 
  - \frac{m_{\drho}^2}{2} \drho_{\mu}^{a} \drho^{a \mu} 
  \label{eqn:rho}\\
& &{} + \frac{1}{2} 
      \left(D_{\mu} \dpi^a \right)^{\dagger} \left(D^{\mu} \dpi^a \right) 
    - \frac{1}{2} m^2_{\dpi} \dpi^a \dpi^a 
    \label{eqn:pi} \\
& &{} - g_{\drho\dpi\dpi} f^{abc} \drho_{\mu}^a \dpi^b D^{\mu} \dpi^c, 
    \label{eqn:lag} 
\end{eqnarray}
It contains the kinetic terms of the vector ($\rho_D$) and 
pseudoscalar ($\pi_D$) mesons, mass terms, and the interactions 
among these mesons.
As we indicated in the introduction, the mesons fill out 
representations of the $SU(2)$ dark flavor symmetry, 
and the meson self-interactions respect the $SU(2)$ dark flavor symmetry.
Throughout all of these expressions, we have assumed that 
the dark sector contains (at least) one set of dark pions and 
(at least) one set of dark vector mesons in the triplet representation 
of the $SU(2)$ dark flavor symmetry.  
Hence the $a = 1,2,3$ index attached to $\pi_D^a$ and 
$\rho_D^a$.\footnote{We use $\rho_D^3$ and $\rho_D^0$ interchangeably.}
We will only consider the phenomenological
consequences of the lightest triplet dark vector meson ($\rho_D^a$) 
and the lightest triplet dark pion ($\pi_D^a$).

The coupling between the 
$\drho$ and $\dpi$ is show in Eq.~\eqref{eqn:lag}. 
This is the analogue of $g_{\rho\pi\pi}$ in QCD\@.
In the $SU(2)_R$ model, the full set of $SU(2)_R$-symmetric
interactions are present, though in practice only the 
$\rho_D^0 \pi_D^+ \pi_D^-$ interaction is phenomenologically 
relevant since only $\rho^0_D$ talks to SM fermions via kinetic mixing (see Sec.~\ref{sec:kinmix}). The NDA estimate of the 
coupling strength is given by
\begin{equation}
g_{\drho \dpi \dpi} \approx \frac{4\pi}{\sqrt{N_D}}.
\label{eqn:rhopipi}
\end{equation}

\subsection{Kinetic Mixing of $\rho_D$ with SM}
\label{sec:kinmix}

The second term of Eq.~\eqref{eq:totlag} contains the kinetic mixing of the dark rhos and 
the electroweak gauge bosons:  
\begin{eqnarray}
\mathcal L_{\rm kinetic \; mixing} \;=\; 
        - \frac{\eps}{2} \drho_{\mu\nu}^{a} F^{a \mu\nu} &=& \left\{ 
   \begin{array}{ll}
   - \frac{\eps}{2} \drho_{\mu\nu}^{a} W^{a \mu\nu} & SU(2)_L \; {\rm model} \\
   - \frac{\eps'}{2} \drho_{\mu\nu}^{0} B^{\mu\nu} & SU(2)_R \; {\rm model}
   \end{array} \right. 
\label{eq:kineticmixing}
\end{eqnarray}
This provides the main ``portal''  
from the Standard Model into the dark sector.  
There are two cases we detail below:  
$F^{a \mu\nu}$ identified with $W^{a \mu\nu}$
(the $SU(2)_L$ model), and 
$F^{a \mu\nu}$ identified with $\delta^{a0} B^{\mu\nu}$
(the $SU(2)_R$ model).

In each of the models defined by Eq.~(\ref{eq:models}), all or part of the 
$SU(2)$ dark flavor symmetry is gauged.  In $SU(2)_L$ model,
the triplet of global $SU(2)$ is identified as a triplet 
of the gauged electroweak $SU(2)_L$ group.  In the $SU(2)_R$ model, 
the triplet of global $SU(2)$ is identified as a triplet 
of the would-be gauged electroweak $SU(2)_R$ group, 
had the entire $SU(2)_R$ been gauged.  Of course the entire
$SU(2)_R$ is not gauged -- just the $U(1)_B$
subgroup.  After electroweak symmetry breaking, 
$SU(2)_L \times U(1)_B \ra U(1)_{\rm em}$, the  
triplet of vector and pseudoscalar mesons of the $SU(2)_L$ and 
$SU(2)_R$ models have the same electric charges, $Q = (+1, 0, -1)$.

In both models, we use  naive dimensional analysis (NDA) to
estimate the size of the kinetic mixing:
\begin{equation}
\begin{array}{ll}
\eps \approx \displaystyle \frac{\sqrt{N_D}}{4\pi} g, & \;\; SU(2)_L \; {\rm model} \\
\eps' \approx \displaystyle \frac{\sqrt{N_D}}{4\pi} g' & \;\; SU(2)_R \; {\rm model} \, ,
\end{array}
\label{eqn:km}
\end{equation}
strictly valid for a large number of colors $N_D$ of the confining 
dark gauge group.
Diagonalizing the kinetic terms leads to a field redefinition of
\begin{equation}
\begin{array}{ll}
W_{\mu}^{a} \ra W_{\mu}^{a} - \eps \, \drho_{\mu}^a & \;\; SU(2)_L \; {\rm model} \\
B_{\mu} \ra B_{\mu} - \eps' \, \drho_{\mu}^0 & \;\; SU(2)_R \; {\rm model} \, ,
\end{array}
\end{equation}
at leading order in $\eps$. This leads to a $\drho$ interaction with 
the SM fermions with a coupling strength proportional to $g^2$
or ${g'}^2$,
\begin{equation}
\Lc_{\drho f \bar{f}} \; = \; \left\{ 
  \begin{array}{ll}
  \eps \,g \,\bar{f}_i \, \bar{\sigma}^{\mu} t^a_{ij} \,\drho_{\mu}^{a} \,f_j & SU(2)_L \; {\rm model} \\
  \eps' Y_f \,g'  \,\bar{f} \, \bar{\sigma}^{\mu} \,\drho_{\mu}^0 \,f & SU(2)_R \; {\rm model} \, ,
  \end{array} \right. 
\label{eq:rhocouplings}
\end{equation}
where $f_{i,j}$ are left-handed SM fermions in the $SU(2)_L$ model,
while $f$ are any SM fermions with hypercharge $Y_f$ in the
$SU(2)_R$ model.

The difference between the two models is mainly in the kinetic mixing.
In the $SU(2)_L$ model, the entire triplet of $\rho_D^a$
mixes with the triplet of $W^a$ bosons.  
In the $SU(2)_R$ model, only the neutral component of the triplet,
$\rho_D^0$, mixes with the hypercharge gauge boson.
Additionally, the kinetic mixing $\eps$ has one power
of the gauge coupling:  $g$ in the $SU(2)_L$ model;
$g'$ in the $SU(2)_R$ model.  Here we emphasize that while the 
difference between $g/g' \simeq 2$ may seem small or trivial, 
$pp \ra \rho$ production is proportional to $3 g^4$ 
in the $SU(2)_L$ model (compared with ${g'}^4$ in the $SU(2)_R$ model), 
and so this leads to a significant difference in the 
production rates of $\rho_D$'s in the two models.\\

Neglecting mass differences among states within the triplets, 
the strong sector is thus described by three parameters:  
\begin{eqnarray}
m_{\pi_D}, \; m_{\rho_D}, \; N_D \qquad {\rm or} \; {\rm equivalently} \qquad 
m_{\pi_D}, \; \eta \equiv \frac{m_{\pi_D}}{m_{\rho_D}}, \; N_D \, .
\end{eqnarray}
As our canonical example that we
use throughout this paper, we have taken $N_D = 4$ in the bulk 
of our results below.  This choice was motivated 
by the Stealth Dark Matter model \cite{Appelquist:2015yfa};
the phenomenology is broadly similar so long as the number of colors
is not excessive.  We quantify this in detail below.

Additionally, we will often replace one of the dark meson mass parameters 
for the ratio $\eta = m_{\dpi}/m_{\drho}$. This ratio is important because 
it governs how the $\drho$ can decay. Specifically, if $\eta < 0.5$, 
$\drho$ can decay to a pair of dark pions, while if $\eta > 0.5$ the dark 
rhos must decay directly back to SM particles. As we will see, the latter case 
is strongly constrained by limits from $Z', W'$ searches. 
From now on, we will label our dark meson models by the type of 
kinetic mixing and the ratio of dark meson masses, i.e., 
\begin{eqnarray}
SU(2)_L^{\eta}:  & & \eps = g \sqrt{N_D}/(4 \pi), \;\; \eps' = 0 \nonumber \\
SU(2)_R^{\eta}:  & & \eps = 0, \;\; \eps' = g' \sqrt{N_D}/(4 \pi) \nonumber 
\end{eqnarray}

\begin{figure}[t]
\includegraphics[width=0.33\linewidth]{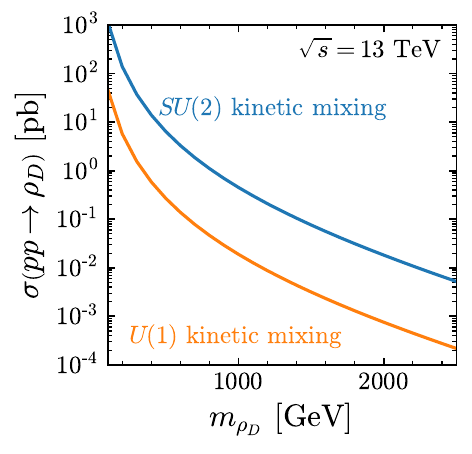}
\includegraphics[width=0.66\linewidth]{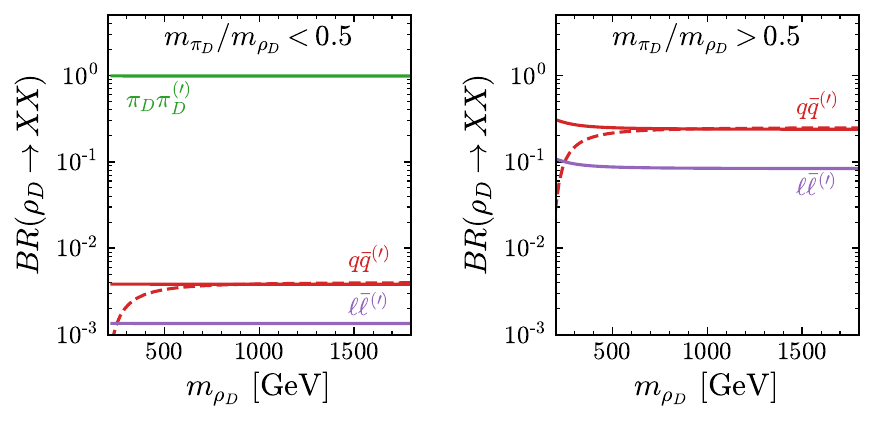}
\caption{The left panel shows the production cross section at 
$\sqrt{s}=13\tev$ for the dark vector mesons. The blue and orange lines 
depict whether the vector mesons are $SU(2)_L$ or $SU(2)_R$ symmetric 
and kinetically mix with the appropriate standard model gauge bosons. 
The middle and right panels show the subsequent branching ratio for 
the $\drho$ depending on whether or not it can decay to the $\dpi$. 
The red lines denote decays to quark anti-quark pairs, and the 
dashed line indicates the top quark. The purple lines show 
leptonic decays.}
\label{fig:RhoBranchingRatio}
\end{figure}

Having specified $N_D$, the production cross section for $\rho_D$ is 
completely determined for both models as shown in the left-side plot in
Fig.~\ref{fig:RhoBranchingRatio}. Figure~\ref{fig:RhoBranchingRatio} 
also shows the $\drho$ branching ratios for two different $\eta$ values: 
as expected, if $\eta < 0.5$ (middle panel) then the interaction strength 
and form of the $\drho \dpi\dpi$ interaction make $\drho \ra \dpi \dpi$. 
On the other hand, if the $\dpi$ are too heavy ($\eta > 0.5$, 
right plot), the $\drho$ decay back through kinetic mixing and the 
branching ratios are simply determined by the SM color factors. 

In focusing on the two models, we are ignoring scenarios where the 
$SU(2)_L \times U(1)_Y$ properties of the $\rho_D$ [and $\pi_D$] 
are not well defined.  Generally, large mixing can only happen in 
scenarios where the strong sector plays a large role in electroweak
breaking and therefore faces constraints from Higgs coupling
measurements and precision electroweak tests. In terms of $\rho_D$
phenomenology, having well defined $SU(2)_L \times U(1)_Y$ properties 
means that the $\rho_D \to V \pi_D$ ($V = $ SM electroweak boson) 
decay modes are always small.

We would be remiss to not point out that the
$SU(2)_R$ model invovling a dark $U(1)$ vector boson 
mixing between the hypercharge is ubiquitious in the
literature of simple dark sectors as ``dark photons'' 
(e.g., for a review \cite{Alexander:2016aln}). 
While most of this literature focuses on (much) lighter
dark photons, for simple dark photon models with a dark photon mass 
at or above the electroweak scale, we can map this toy model 
onto a special case of our strongly-coupled dark sector.  
The mapping utilizes the $SU(2)_R$ model with: 
$\eta > 0.5$ (so that the dark vector boson can decay 
only into SM states), 
$m_{\rm SM}/v_\pi$ small 
(so that single production of dark pions is negligible),
and the number of dark colors $N_D$ chosen to obtain
a kinetic mixing $\epsilon'$.  Even with these parameter choices,
our strongly-coupled dark sector obviously has differences 
from the simple toy models.  One is that the kinetic mixing is 
at most one-loop suppressed.  Another is that there is 
relationship between the smallness of the kinetic mixing,
the number of dark colors, and the relative size of 
self-interactions of the dark mesons.  While it would be 
interesting to map out this space more fully, this is 
beyond the scope of this paper.

\subsection{Dark Pion Decay to SM}

Finally, dark pion decay.  This is the main subject of our 
companion paper \cite{Kribs:2018oad}.  There we show that strongly-coupled 
models with custodially-symmetric Higgs interactions among the dark 
fermions leads to a low energy effective theory in which dark pions 
interact with the SM through:
\begin{eqnarray} 
\mathcal L_{decay} &=& \frac{\sqrt{2}}{v_\pi} \, 
\bigg[ \, \dpi^+ \bar{\psi}_u (m_d P_R - m_u P_L) \psi_d \, 
+ \, \dpi^- \bar{\psi}_d (m_d P_L - m_u P_R) \psi_u \nonumber \\
& &{} \qquad\qquad\quad + \frac{i}{\sqrt{2}} \, \dpi^0 (m_u \, \bar{\psi}_u \gamma_5 \psi_u - m_d \, \bar{\psi}_d \gamma_5 \psi_d) \, \bigg]  \nonumber \\
& & - \, \xi \, \frac{m_W}{v_\pi} 
\left[ (W_\mu^-\, h\overleftrightarrow \partial^\mu \dpi^+) + (W_\mu^+\, h \overleftrightarrow \partial^\mu \dpi^-) 
+ \frac{1}{\cos{\theta_W}} (Z_\mu\, h \overleftrightarrow \partial^\mu \dpi^0) \right] 
\label{eq:ldecay}
\end{eqnarray} 
where $\psi_{u,d}$ are SM fermions.  There are several important features
of this Lagrangian. First, while we have used the language that the decay interactions `break the flavor symmetry', this is slightly sloppy. Stated more correctly, we have married the $SU(2)_V$ symmetry of the dark pions to part of the $O(4)$ symmetry group of the Higgs potential. Both the dark pions and the SM fields transform under the shared symmetry, so we can write down single pion interactions of the form $\pi^a \mathcal O^a$ where $\mathcal O^a$ is some triplet of SM fields.

 The overall scale of the operators
is set by $1/v_\pi$ for the fermions and $\xi/v_\pi$ for the 
gauge/Higgs bosons.  The fact that the interactions do not further
distinguish the fermions (i.e., one overall coupling for the 
first four terms) nor the gauge/Higgs interactions
(one coupling for the last three terms)
is due to the the dark sector's preservation of custodial symmetry.  
However, since custodial $SU(2)$ is broken 
in the SM by differences of Yukawa couplings as well as hypercharge, 
there is a residual differentiation of the interactions by
$m_u - m_d$ as well as $g' \not= 0$.  
 
This form is convenient, since coupling $\dpi$ to the SM fields 
requires breaking electroweak symmetry and hence the 
coupling strengths must be proportional to the mass of a SM 
field. The primary role of the $1/v_\pi$ parameter is to set the total width 
of the $\dpi$.  In this paper our main focus is on scenarios where the 
$\dpi$ decay promptly. This sets a lower bound on $m_{SM}/v_\pi$, 
where $m_{SM}$ is the mass of the mass of the SM particle(s) in 
the dominant $\dpi$ decay.  Scenarios where $\dpi$ is displaced 
or long-lived are also interesting to study.  The main search
methodologies are well-known from other displaced/long-lived 
searches (for a review, see e.g.~\cite{Lee:2018pag}).

The remaining model-dependent parameter is the relative strength of 
the coupling to fermions versus the gauge/Higgs sector that 
we have parameterized by $\xi$.  We will consider two possibilities
for $\xi$:
\begin{equation}
\boxed{
\begin{array}{rclcc}
\xi & = &  1 \;\; & \mbox{``gaugephilic''} \\
\xi & = &  c_\xi \displaystyle \frac{v^2}{m_{\pi_D}^2} \ll 1 \;\; & \mbox{``gaugephobic''} 
\end{array}}
\label{eq:gaugephobicgaugephilic}
\end{equation}
The scaling of the gaugephobic parameter with the electroweak scale
and the dark pion mass scale deserves some discussion.
The origin of this scaling is found from an analysis of the 
strongly-coupled effective theories that we have discussed in detail in
Ref.~\cite{Kribs:2018oad}.  In essence, there are higher dimensional
operators involving additional Higgs fields, suppressed by at least
the scale of the dark pions, that can regenerate couplings to
the gauge/Higgs sector even if they don't exist at leading order.
As we show in Ref.~\cite{Kribs:2018oad}, the Stealth Dark Matter model
is gaugephobic with $\xi = m_h^2/(m_{K_D}^2 - m_h^2) \simeq m_h^2/m_{K_D}^2$
where $K_D$ is a another dark pion that is at least slightly heavier than 
$\pi_D$.  Since the dark kaon scales with the parameters of 
the ultraviolet theory in exactly the same way as the dark pion,
in our phenomenological study we take $c_\xi = \lambda_h$
and do not distinguish between the dark pion and kaon masses.

In the limit that the dark pion mass scale is taken large, 
$\xi \ra 0$, and the dark pions can only decay back to fermions.
However, when the dark pions are near to the electroweak scale, 
$\xi$ can be ``smallish'' but, importantly, nonzero.  
This implies $\pi_D \rightarrow f \bar{f}'$ dominate 
so long as there is no small coupling.  For the specific case
of $\pi_D^0$ in the mass range $m_h + m_Z < m_{\pi_D^0} < 2 m_t$,
the decay $\pi_D^0 \ra Z + h$ dominates despite 
being gaugephobic.  This is because the $Z h$ mode
is longitudinally enhanced, while the competing fermionic mode 
$\pi_D^0 \ra b \bar{b}$ is suppressed by the small 
Yukawa coupling $y_b$.
For all other ranges of dark pion masses (both charged and neutral),
$\pi_D \rightarrow f \bar{f}'$ dominates.  By contrast, in the 
gaugephilic case $\pi_D \ra W + h, \, Z + h$ dominate once 
they are kinematically open. 

While the two choices in Eq.~(\ref{eq:gaugephobicgaugephilic}) 
may seem arbitrary at first, 
a large class of strongly-coupled models can be mapped into 
this categorization (see 
Ref.~\cite{Kribs:2018oad} for more details). Specifically, the Stealth Dark 
Matter model~\cite{Appelquist:2015yfa,Appelquist:2015zfa,Kribs:2016cew} 
and others similar to it are gaugephobic.  By contrast, models
of bosonic technicolor / induced symmetry breaking~\cite{Chang:2014ida},
as well as the triplet state in Georgi-Machacek models~\cite{Gunion:1989we} 
have gaugephilic interactions.  

\begin{figure}[t]
\begin{center}
\includegraphics[width=0.95\linewidth]{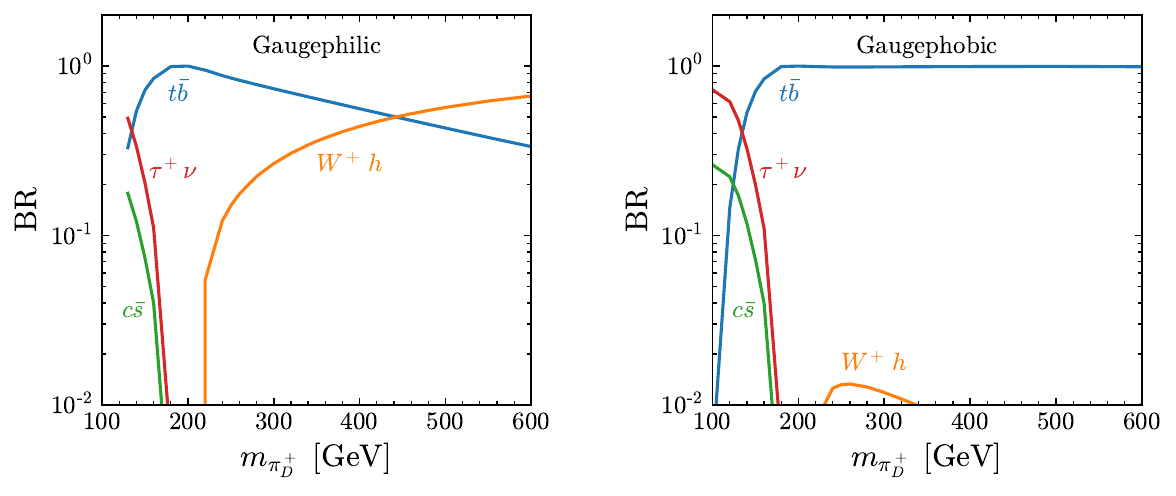}
\end{center}
\caption{Branching ratios of the charged pions}
\label{fig:ChargedPionBranchingRatio}
\end{figure}

In our taxonomy, the gaugephilic case only occurs for the $SU(2)_L$ model.  
This is not immediately obvious from our discussion thus far.
Essentially the gauge/Higgs interactions on the last line
of Eq.~(\ref{eq:ldecay}) is permitted with order one $\xi$
when $\pi_D^a$ is in the same representation as $W_\mu^a$,
i.e., an $SU(2)_L$ triplet.  The reader may then immediately wonder 
why the $SU(2)_R$ case does not have $\xi = 0$.  At leading order
it does, but at higher orders one finds gauge/Higgs interactions 
are generated albeit with a suppression typically of order 
$m_h^2/m_{\pi_D}^2$.  This is parametrically the suppression we find
in the Stealth Dark Matter model \cite{Kribs:2018oad}, and is 
similar to what we find in generic 2-flavor custodially-symmetric 
models.  More details can be found in Ref.~\cite{Kribs:2018oad}.

Any given model may or may not permit arbitrary choices for
$v_\pi$ and $\xi$; for instance, induced electroweak symmetry breaking
requires $v_\pi$ fixed (up to order one coefficients) and $\xi = 1$
due to the requirements of proper electroweak symmetry breaking. 
However, as we detail in~\cite{Kribs:2018oad}, there are 
models that span a wide range of $(v_\pi, \xi \lesssim 1)$.

\begin{figure}[t]
\begin{center}
\includegraphics[width=0.95\linewidth]{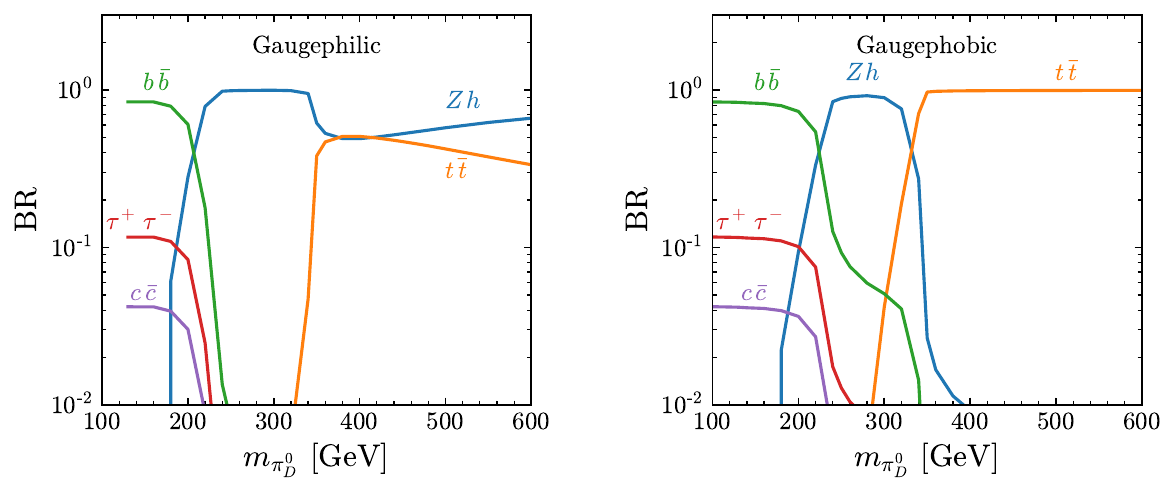}
\end{center}
\caption{Branching ratios of the neutral pions}
\label{fig:NeutralPionBranchingRatio}
\end{figure}

Given $\xi$, the branching fractions of the $\dpi$ are fully specified as 
a function of the pion mass. As $\dpi$ decay couplings are proportional 
to mass, they decay to the heaviest kinematically available SM particles. 
The branching ratios for the gaugephilic and gaugephobic scenarios 
are compared side by side in 
Fig~\ref{fig:ChargedPionBranchingRatio} (charged $\dpi$) 
and \ref{fig:NeutralPionBranchingRatio} (neutral $\dpi$).\footnote{We have omitted the 
anomaly-induced decay $\dpi^0 \to \gamma\gamma$ from 
Fig.~\ref{fig:NeutralPionBranchingRatio}.  In models with a 
$SU(2)$ flavor symmetry that becomes custodial $SU(2)$ after
Higgs interactions, the dark sector is anomaly-free. 
The decay mode does reappear due to SM interactions violating
custodial $SU(2)$, but is highly suppressed so as to be
phenomenologically irrelevant \cite{Kribs:2018oad}.}

For the charged $\dpi$, the branching ratios in the two cases are similar 
at small masses. However, the unsuppressed gauge/Higgs couplings in the 
gaugephilic scenario imply $\dpi \ra W^+\,h$ quickly dominates 
once it is kinematically allowed (due to the kinematic enhancement of 
decays to longitudinal $W$), while the $\dpi \rightarrow t \bar{b}$ mode 
always dominates at heavy mass for the gaugephobic case. There is a 
similar pattern in the branching ratio of the neutral pions. Again, when 
the pion is light, the decay modes between the two categories are similar 
and are dominated by the $b\bar{b}$ mode. This similarity persists after 
$\dpi$ passes the $Z h$ threshold. However, as $\dpi$ is further increased 
past the $t\bar{t}$ threshold we can spot the difference, as the $\dpi \to 
\bar t t$ branching ratio dominates at large $\dpi$ masses in the 
gaugephobic case but stays subdominant to $Z h$ in the gaugephilic case.

\section{Constraints from single production}
\label{sec:collider}

Having established the dark meson phenomenological Lagrangian and fleshed 
out the relevant parameters, we now move on to LHC production, 
sensitivities, and constraints.

The phenomenology of the dark meson sector that we pursue in this
paper clearly bifurcates at $\eta = 0.5$ as evident from 
the branching fractions of the dark pions in 
Fig.~\ref{fig:RhoBranchingRatio}. 
For $\eta > 0.5$, 
the $\drho$ is kinematically forbidden to decay to a pair of 
on-shell dark pions, and thus decays to SM fermions 
dominate.\footnote{Fig.~\ref{fig:RhoBranchingRatio} includes three-body 
decays though an off-shell dark pion, but the rates for these decay modes
are always small compared to what is shown in the figure.}  The decays into 
SM fermions are determined solely by the gauge and color charges of the 
fermions, so the $\drho$ phenomenology is essentially independent of the 
details of how the pions interact with the SM\@.

When $\eta < 0.5$, $\drho \to \dpi\dpi$ is open, and generally
dominates so long as the number of dark colors, $N_D$, is not 
large (we'll be more precise below).  In this case, the most promising
way to search for dark mesons is dark pion pair production.  
The largest contribution to dark pion pair production is 
resonant production $pp \to \drho \to 
\dpi\dpi$ through the dark rho, so long as it is not very heavy.
Dark pions can also be pair-produced through Drell-Yan production,
though this tends to give a smaller cross section due to the 
$W$ or $Z$ exchange being off-shell.  We find that resonant 
production through $\drho$ dominates for $\eta \gtrsim 0.2$
for $N_D = 4$.

The final states populated by dark pion pairs depends on how 
the dark pions decay, which in turn depends on whether we 
are in a gaugephilic or gaugephobic scenario.
We have chosen 9 benchmarks spanning the phenomenology possibilities
that we believe give a solid idea of the differing phenomenology,
shown in Table~\ref{table:benchmarks}.
We provide the \texttt{FeynRules}~\cite{Alloul:2013bka} model files 
and corresponding UFO files on
\href{https://github.com/bostdiek/HeavyDarkMesons}{GitHub}.\footnote{ https://github.com/bostdiek/HeavyDarkMesons}

We used 
\texttt{MadGraph5\_aMC@NLO}~\cite{Alwall:2014hca} to simulate the events. 
When studying constraints directly on the $\drho$, we simulated 
$pp\rightarrow \drho$ and then allowed for any decay mode. For the 
constraints on $\dpi$, we simulated $pp\rightarrow \dpi \dpi$ which then 
had resonant and Drell-Yan production. In all cases, showering and 
hadronization was performed by \texttt{Pythia 8}~\cite{Sjostrand:2014zea} 
and \texttt{Delphes 3}~\cite{deFavereau:2013fsa} was used for fast 
detector simulation. We used the default detector card because we recast 
both ATLAS and CMS results. Within \texttt{Delphes}, jets were calculated 
with \texttt{FastJet}~\cite{Cacciari:2011ma} using the anti-$k_t$ 
algorithm \cite{Cacciari:2008gp}

For each of the benchmark scenarios in Table~\ref{table:benchmarks}, the mass 
of the $\dpi$ was scanned with variable spacing in 
order to capture the different decay mode transitions. We take the lower 
limit of dark pion mass to be $100$~GeV, coming from the bound on 
BSM charged particles from LEP II\@.
At each mass point, 
500k events were produced for pair production of dark pions (all allowable 
modes). This was done for both $\sqrt{s}=8\tev$ and $\sqrt{s}=13\tev$ 
collisions. The $\dpi$ are decayed in the narrow width approximation using 
\texttt{Pythia}.

There is no dedicated search for dark mesons at the LHC\@.  
We therefore estimate the existing bounds by recasting a vast set of
potentially constraining searches using Monte Carlo methods. We will 
present our results first, followed by a more detailed description of our 
recasting methods and a summary of why several searches which look 
promising at first glance fail to set strong bounds. 

\begin{table}[t] 
\renewcommand{\arraystretch}{1.2}
\setlength{\tabcolsep}{5pt}
\setlength{\arrayrulewidth}{.3mm}
\begin{center} 
\begin{tabular}{c|cl} \hline\hline
Model     & $\eta \equiv m_{\pi_D}/m_{\rho_D}$ & $\qquad \xi$ \\ \hline 
$SU(2)_L^{55}$ & $0.55$ & \\ 
$SU(2)_L^{45}$ & $0.45$ & gaugephilic ($\xi = 1$) \\ 
$SU(2)_L^{25}$ & $0.25$ & \\ \hline
$SU(2)_L^{55}$ & $0.55$ & \\ 
$SU(2)_L^{45}$ & $0.45$ & gaugephobic ($\xi = m_h^2/m_{\pi_D}^2$) \\ 
$SU(2)_L^{25}$ & $0.25$ & \\ \hline
$SU(2)_R^{55}$ & $0.55$ & \\ 
$SU(2)_R^{45}$ & $0.45$ & gaugephobic ($\xi = m_h^2/m_{\pi_D}^2$) \\ 
$SU(2)_R^{25}$ & $0.25$ & \\ \hline\hline
\end{tabular} 
\end{center} 
\caption{Benchmark models and parameters used in our study.
Note that the gaugephilic case only occurs for the 
$SU(2)_L$ model, as discussed in Sec.~\ref{sec:phenodarkmesons} 
in the text.}
\label{table:benchmarks}
\end{table} 

\subsection{$\rho_D$ constraints}
\label{sec:rhoD}

We first consider $\rho_D$ production and decay.  The 
$\rho_D$ dark vector mesons kinetically mix with electroweak gauge
bosons, shown in Eq.~(\ref{eq:kineticmixing}), giving direct
couplings to SM fermions, shown in Eq.~(\ref{eq:rhocouplings}).
In both the $SU(2)_L$ and $SU(2)_R$ models, there is a neutral
$\rho_D^0$, better known as a new $Z'$ gauge boson. Via kinetic mixing, this $\rho_D^0$ acquires a coupling to leptons.

The strongest constraints on generic $Z'$ gauge bosons (with 
masses near or above the electroweak scale) is from the absence of 
resonances in the the $\ell^+ \ell^{-}$ invariant mass spectrum 
\cite{Aaboud:2017buh, Sirunyan:2018exx}.  
Using the ATLAS 13 TeV search with 36.1 fb$^{-1}$ of integrated 
luminosity \cite{Aaboud:2017buh}, we have recast the dilepton searches
for the combined electron and muon channels into a limit on 
$\drho$ cross section times branching fraction to leptons.
This is accomplished by simulating the production of $\drho$ 
and decaying them according to the branching ratios shown 
in Fig~\ref{fig:RhoBranchingRatio}. 
After passing through a parton shower, hadronization, and detector 
simulation, we select events which contain same-flavor opposite-sign 
leptons within the ATLAS selection criteria. 
The combined efficiency (branching ratio times the detector efficiencies)
multiplied by the cross section can then be compared against 
the exclusion limits provided by the ATLAS 
HEPData~\cite{HEPDATA_DILEPTON}. 

\begin{figure}[t]
\begin{center}
\includegraphics[width=0.99\linewidth]{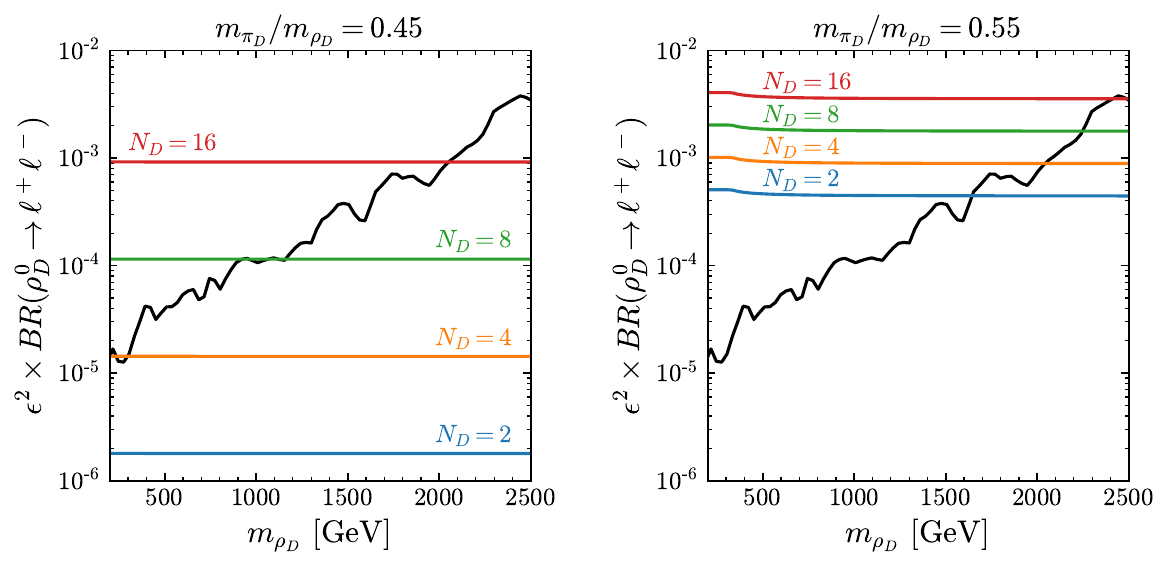}
\end{center}
\caption{Constraints on the kinetic mixing 
between the the SM and $\rho^0_D$ 
(times the leptonic branching fraction of $\rho^0_D$) 
from the non-observation of a dilepton resonance near $m_{\rho_D}$.
The black line is the model-independent limit.
To illustrate the impact of this bound on the model space,
we have superimposed the predicted 
$\eps^2 \times {\rm BR}(\rho^0_D \ra \ell^+\ell^-)$ 
for the $SU(2)_L$ model, varying the number of colors
between $2$ to $16$.  On the right, 
the 2-body decay 
$\rho_D^0 \ra \pi_d^+ \pi_d^-$ is kinematically forbidden,  
leading to strong constraints: $m_{\rho_D} > 1.5$-$2.5$~TeV\@.
On the left, the 2-body decay 
$\rho_D^0 \ra \pi_d^+ \pi_d^-$ is open, 
and we see that when $N_D \lsim 4$, there is no constraint
from resonant $\rho^0_D$ production and decay to dileptons.}
\label{fig:Dileptonlimit}
\end{figure}

In Fig.~\ref{fig:Dileptonlimit}, 
we illustrate the bounds that we have obtained by determining the
largest coupling of the $\rho_D^0$ to the SM for any choice of
$m_{\rho_D}$ within the range of interest in this paper.
The coupling is completely determined by the model-independent 
quantity $\eps^2 \times BR(\rho_D^0 \ra \ell^+\ell^-)$,
that is shown as a black line in both panels of Fig.~\ref{fig:Dileptonlimit}.
Also superimposed on the panels are the predicted sizes of
$\eps^2 \times BR(\rho_D^0 \ra \ell^+\ell^-)$ for a given
$m_{\pi_D}/m_{\rho_D}$ and number of dark colors $N_D$
in the $SU(2)_L$ model.  It is important to note that $\eps$ is the kinetic 
mixing parameter and not the detector efficiency. (Similar but weaker constraints
are found in the $SU(2)_R$ model.)
The right panel clearly shows that the neutral dark vector meson 
is strongly constrained by the dilepton data when 
$m_{\pi_D}/m_{\rho_D} > 0.5$.  

The dependence on 
the number of dark colors is nontrivial:
\begin{eqnarray}
\sigma( p p \ra \rho_D^0 \ra \ell^+ \ell^-) \; \propto \; 
\eps^2 \times BR(\rho_D^0 \ra \ell^+ \ell^-) \propto
\left\{ 
\begin{array}{ll}
N_D   & \quad \eta > 0.5 \\
N_D^3 & \quad \eta < 0.5 \, .
\end{array}
\right.
\end{eqnarray}
In the case $\eta > 0.5$, the one power of $N_D$ comes from 
$\eps^2$ while in the branching fraction the $N_D$ dependence
cancels.  Contrast this with the case $\eta < 0.5$,
where the branching fraction
$BR(\rho_D^0 \ra \ell^+ \ell^-)|_{\rm \eta < 0.5} \simeq
\Gamma(\rho_D^0 \ra \ell^+ \ell^-)/\Gamma(\rho_D^0 \ra \pi_D \pi_D) 
\propto N_D^2$.  The left panel clearly shows that when 
$\rho_D \ra \pi_D \pi_D$ is both kinematically open 
($\eta < 0.5$) and dominates ($N_D \lesssim 4$), 
there are virtually no LHC constraints on neutral dark vector 
meson production and decay.  (The very narrow region near 
$m_{\rho_D} \sim 300$~GeV is, as we will see, also 
constrained by other searches). 

The bounds we have obtained from the ATLAS searches for dilepton 
resonances assumed the width of the new resonance is 
relatively narrow, $\Gamma(Z')/M_{Z'} \lesssim 0.03$ \cite{Aaboud:2017buh}.
In all of the cases with $\eta = 0.55$,
where the $\rho_D^0$ can only two-decay into SM states,
the width is narrow, 
$\Gamma_{\rm tot}(\rho_D^0)/m_{\rho_D^0} < 10^{-3}$.  
Once $\rho_D \ra \pi_D \pi_D$ is open, we can estimate 
this partial width \cite{Kilic:2009mi}
\begin{eqnarray}
\frac{\Gamma(\rho_D \ra \pi_D \pi_D)}{m_{\rho_D}} 
  &=& \frac{\pi}{3 N_D} 
      \left( 1 - \frac{4 m_{\pi_D}^2}{m_{\rho_D}^2} \right)^{3/2}
  \simeq 
    \frac{4}{N_D} \times \left\{ \begin{array}{ll}
       0.02 & \;\; \eta = 0.45 \\
       0.16 & \;\; \eta = 0.25  \, ,
    \end{array} \right. 
\end{eqnarray}
where we have evaluated the result for the two values $\eta = 0.45, \, 0.25$
used in our benchmarks for the paper.  
Despite the relative strong-coupling among
mesons ($g_{\rho_D \pi_D \pi_D} = 4 \pi/\sqrt{N_D} \gg 1$), 
the kinematic suppression of taking $\eta = 0.45$ suppresses the width 
of the $\rho_D^0$ to a few percent, and so the ATLAS bounds are fully
applicable.  For $\eta = 0.25$, the width is now tens of percent
that is large enough requiring a re-analysis of the dilepton data 
to set precise bounds on the $\rho_D^0$.  For $\eta = 0.25, N < 4$, the
$\rho_D$ width to mass ratio reaches $\sim$ tens of percent, 
so a simple recast of the ATLAS bounds is not completely precise. 
However, given that i.) the bounds on a wide resonance will be weaker 
than on a narrow resonance, and ii.) the narrow resonance bounds for 
$N<4$ are already weak, we conclude that there is no bound on 
$\rho^0_D$ for $\eta = 0.25, N < 4$.

There is one additional constraint on the kinetic mixing of 
$\rho_D$ with SM gauge bosons from LEP constraints on four-fermion
effective operators \cite{Schael:2013ita}.
Integrating out $\rho_D^0$ results in
four-fermion operators of the form
\begin{equation}
\frac{4 \pi}{\Lambda^2} \, \bar{e} e \bar{f} f \, ,
\end{equation}
where we have used the operator normalization of Ref.~\cite{Schael:2013ita}.
Matching the coefficient,
\begin{eqnarray}
\frac{4 \pi}{\Lambda^2} &=& 
\frac{1}{m_{\rho_D}^2} \times  
  \left\{ \begin{array}{cc} \eps^2 g^2        & \quad SU(2)_L \; {\rm model} \\ 
                            {\eps'}^2 {g'}^2  & \quad SU(2)_R \; {\rm model}  
                            \end{array} 
  \right. \nonumber \\
 &=&{} \frac{N_D}{16 \pi^2 m_{\rho_D}^2} \times \left\{ 
  \begin{array}{cc} g^4 & \quad SU(2)_L \; {\rm model} \\ 
                    {g'}^4 & \quad  SU(2)_R \; {\rm model}  
  \end{array} \right. \, .
\end{eqnarray}
The strongest constraints from the LEP data suggest $\Lambda \gtrsim 20$~TeV 
\cite{Schael:2013ita}.  For the $SU(2)_R$ model, there is
no constraint due to the smallness of $g'$.  
For the $SU(2)_L$ model, the bound on $m_{\rho_D^0}$
varies from about $250$--$750$~GeV for $N_D = 2$--$16$.
Given the order one uncertainties in the large $N_D$ estimate for
the kinetic mixing, this bound is not any stronger than what we 
have already found from Fig.~\ref{fig:Dileptonlimit}.

\subsection{Constraints on the dark pion coupling to SM}

Throughout the paper, we will generally work in the ``vector-like''
limit (See Ref.~\cite{Kribs:2018oad}) where $\frac{m_{SM}}{v_\pi}$ 
is small and thus single production 
of $\pi_D$ is suppressed.  This limit is automatically safe from 
constraints from electroweak precision observables as well as 
Higgs coupling measurements, and coincides with the demarcation
of our model space into the two categories $SU(2)_L$ and $SU(2)_R$.  
If, however, $\frac{m_{SM}}{v_\pi}$ is not so small,
single production of dark pions is possible and relevant to 
the phenomenology.  In the model of 
bosonic technicolor / induced electroweak symmetry breaking, 
this sets the strongest constraints \cite{Chang:2014ida}.

\begin{figure}[t!]
\begin{center}
\includegraphics[width=0.5\linewidth]{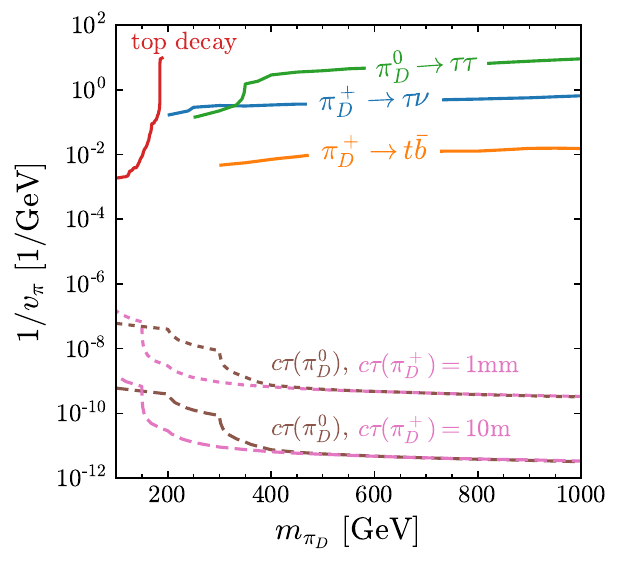}
\end{center}
\caption{Constraints on the value of $1/v_{\pi}$ as a function of 
the dark pion mass. Precise measurements of the top quark exclude regions 
above the red line. The green, blue, and orange lines come from collider 
searches for heavy Higgs particles (mainly in 2HDM). Lastly, the brown and 
pink dashed lines are not constraints, but show at what point the 
phenomenology changes. Below these lines, the pions start to travel an 
appreciable distance in the detector, either leading to displaced vertices 
or disappearing tracks. The lower of these lines are around the scale when 
the particles leave the detector either as missing energy or look like 
stable charged particles.}
\label{fig:singlepi}
\end{figure}

We can also characterize the parameter space of our effective theory
by determining the constraints on $1/v_\pi$ of Eq.~(\ref{eq:ldecay}).  In Fig.~\ref{fig:singlepi},
we consider several processes\footnote{Note 
that we only consider processes involving fermions  
so that we have $\xi$-independent constraints on $1/v_\pi$.
Larger $\xi$, e.g., $\xi \sim 1$, there can be stronger
constraints from couplings to the gauge/Higgs sector
\cite{Chang:2014ida}.  We thank Ennio Salvioni for 
discussions on this point.}
where single dark pion production
can set upper bounds on $1/v_\pi$.  
One process is top decay,
$t \ra \pi^+_D \bar{b}$.  In this process the $\pi^+_D$ must be 
somewhat lighter than the top quark, and thus 
$\pi^+_D \ra \tau^+ \nu_\tau$ dominates for the charged pions, leading 
to an excess of $\tau$'s in top decay.
LHC analyses of top decay, however, are 
consistent with lepton universality \cite{Aaltonen:2014hua, Aad:2015dya}. 
For values of the pion mass slightly less 
than the top quark mass, the pion branching ratio to $\tau$ is 
similar to the SM branching ratio of the $W$ to tau. 
Thus, in this region 
the branching ratio alone is not enough to constrain the coupling. 
Instead, we use the total width of the top quark 
\cite{Tanabashi:2018oca,Abazov:2012vd,Khachatryan:2014nda} as a 
secondary constraint, and exclude any region where the BSM additions to 
the top decay change either the width or the tauonic branching ratio by 
more than two standard deviations away from the measured values.
This constraint is shown in red in Fig.~\ref{fig:singlepi}.

There are also many searches for the heavy Higgs particles of 
two-Higgs doublet models that can be recast into searches
for single production of the charged or neutral dark pions.
In Refs.~\cite{ATLAS:2016grc,ATLAS:2016qiq}, ATLAS searches for a 
charged Higgs produced  association with $t\bar{b}$. 
The two searches consist of one looking for 
$H^+\rightarrow \tau^+ \nu_{\tau}$ while the other
looks for $H^+ \rightarrow t \bar{b}$.  
The limits are presented in terms of $\sigma(t\bar{b} H^+) \times BR$, 
but unfortunately HEPData is not given.  We therefore take the limits 
from plots in Refs.~\cite{ATLAS:2016grc,ATLAS:2016qiq} and 
reinterpret them by replacing $\pi_D^+$ for the charged Higgs boson. 
The upper bounds on $1/v_\pi$ we obtain are shown in orange and blue 
in Fig.~\ref{fig:singlepi}.  Finally, in a similar approach, 
Ref.~\cite{Aaboud:2017sjh} performed searches for a heavy 
neutral Higgs boson produced in association with $b\bar{b}$ and decaying to $\tau^+\tau^-$.
Upon recasting this search for neutral dark pions, we find 
somewhat weaker constraints -- shown in green in Fig.~\ref{fig:singlepi} -- compared with the bounds from charged dark pions.

Finally, while this is not a constraint on the parameter space
per se, it is interesting to determine when $1/v_\pi$ is 
small enough that the decays of the dark pions are no longer 
prompt in colliders.  As a rough guide, we can use
\begin{equation}
\Gamma = \left(\frac{2~\rm{ mm}}{c\tau} \right) \times 10^{-13} ~\rm{ GeV}
\end{equation}
and estimate that if $c\tau = 1$~mm, then the neutral pions would lead to 
displaced tracks, or the charged pions would lead to kinked (or 
disappearing) tracks when they decay. If $c\tau > 10$~m, then the pions 
can escape the detectors before decaying, leading to missing energy 
or long-lived charged tracks. 
Search strategies for both of these types 
of signals are interesting but best explored through existing dedicated
strategies for long-lived charged or neutral particles~\cite{Liu:2015bma,Evans:2016zau, Curtin:2018mvb,Liu:2018wte}.  
The smallness that $1/v_\pi$ needs to be to lead to 
these long-lived signals is shown in Fig.~\ref{fig:singlepi}.

There can also be a contribution to the $S$ parameter as a result
of the interactions in Eq.~(\ref{eq:ldecay}).  However, in the 
ultraviolet strongly-coupled theories considered in Ref.~\cite{Kribs:2018oad}, 
we find the contributions depend on the spectrum of the 
heavier mesons, and so there is no useful translation into
bounds on $1/v_\pi$.  Suffice to say that there are no
bounds from the $S$ parameter when the contributions to the 
dark fermion masses are mostly vector-like with only smaller 
contributions arising from electroweak symmetry breaking
\cite{Carone:1992rh,Barbieri:2007bh}. 

Clearly, there is a huge range in $1/v_\pi$ -- roughly values larger than $10^{-7}$ and smaller than $10^{-2}$, with some slight variation depending on $m_{\pi_D}$ -- where dark pion decays
are prompt but the rate for single dark pion production is too small
to be detected. Our goal for the remainder of this paper is to explore how prompt LHC searches constrain paired dark pion production in this otherwise open region of parameter space.

\section{Resonant Dark Pion Pair-Production at LHC}
\label{sec:Resonant}

The rate for dark pion pair-production depends on the 
model -- $SU(2)_L$ versus $SU(2)_R$, and the NDA estimates 
for the kinetic mixing as well as the meson self-interactions.
It does \emph{not} depend on how the dark pions decay
(gaugephilic versus gaugephobic) because the production
rate is independent of $1/v_\pi$ and $\xi/v_\pi$ from
Eq.~(\ref{eq:ldecay}). 
However, the different decay 
modes require different search strategies. In Table~\ref{tab:Regions}, 
we have denoted different mass regions for each of the categories defined 
by which decay modes are dominant. The intermediate SM particles, 
which may subsequently decay, are listed for both the charged 
$\left(\dpi^{\pm}\,\dpi^0\right)$ and the neutral 
$\left(\dpi^+ \dpi^- \right)$ currents. Note that the symmetries 
do not allow for neutral currents of the type $\dpi^0 \dpi^0$, so the $SU(2)_R$ model does not contain a resonantly enhanced charged current. 
\begin{table}[t!]
\centering
\renewcommand{\arraystretch}{1.7}
\setlength{\tabcolsep}{5pt}
\setlength{\arrayrulewidth}{.3mm}
\begin{tabular}{c |  c | c c }
\hline
\hline
&  Mass& Charged Current & Neutral Current \\
\hline
\multirow{4}{*}{\rotatebox{90}{gaugephilic}} &  $m_{\pi_D} \lesssim 150 \gev$ & $b\bar{b} \tau \nu$ & $ \tau^+ \tau^- \nu \bar{\nu}$ \\
 &  $150 \gev \lesssim m_{\pi_D} \lesssim 200 \gev$ & $b\bar{b} t\bar{b}$ & $t\bar{t} b\bar{b}$  \\
 &  $200 \gev \lesssim m_{\pi_D} \lesssim 450 \gev$ & $Z\, h\,t\bar{b}$ & $t\bar{t} b\bar{b}$ \\
 &  $m_{\pi_D} \gtrsim 450 \gev$ & $h\,h\,Z\,W^+$ & $hhW^+W^-$ \\
\hline
\hline
\multirow{4}{*}{\rotatebox{90}{gaugephobic}}& $m_{\pi_D} \lesssim 150 \gev$ & 
$b\bar{b} \tau \nu$ & $ \tau^+ \tau^- \nu \bar{\nu}$ \\
& $150 \gev \lesssim m_{\pi_D} \lesssim 220 \gev$ & 
$b\bar{b} t\bar{b}$ & $t\bar{t} b\bar{b}$  \\
& $220 \gev \lesssim m_{\pi_D} \lesssim 350 \gev$ & 
$Zht\bar{b}$ & $t\bar{t} b\bar{b}$ \\
& $m_{\pi_D} \gtrsim 350 \gev$ & 
$t\bar{t} t\bar{b}$ & $t\bar{t} b\bar{b} $\\
\hline
\hline
\end{tabular}
\caption{Phenomenological regions for collider signatures. The charged and neutral current columns show the SM particles for the dominant branching ratios.}
\label{tab:Regions}
\end{table}

Table~\ref{tab:Regions} shows that there are many Standard 
Model particles in the final states, with possibly exotic combinations. 
We analyzed 13 searches (in addition to the ones already discussed), 
broken down into 6 searches at $8$~TeV and 7 searches at $13$~TeV\@. 
Surprisingly, we find that many of the searches
are not sensitive to our benchmark models. The searches with sensitivity are 
further detailed here, while we save a discussion of the non sensitive searches
for Sec.~\ref{sec:Additional}.

\begin{figure}[t!]
\begin{center}
\includegraphics[width=\linewidth]{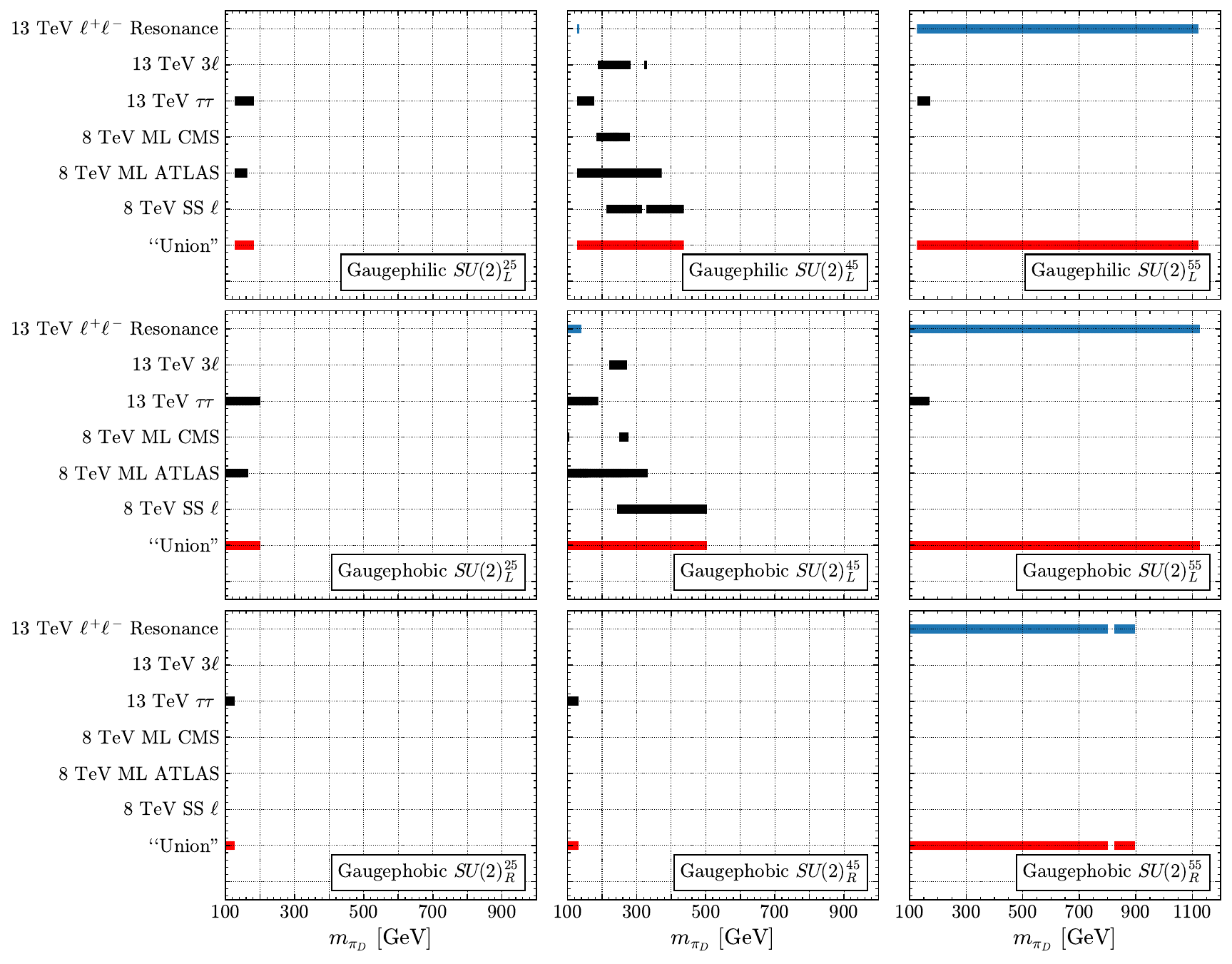}
\end{center}
\caption{Summary of the dark meson exclusions for the benchmark scenarios and values of the $\pi_D$ and $\rho_D$ masses. The scenarios are labeled by the type of kinetic mixing, the ratio of the dark pion to dark rho mass $\eta = m_{\pi_D}/m_{\rho_D}$, and the relative strength of the fermionic versus bosonic dark pion decay modes. All of the dark pions decay promptly. The top line indicates the bound on $\rho^0_D$ inferred from recasting the latest dilepton bounds and interpreted in terms of $m_{\pi_D}$. The next five lines (in black) show the $\pi_D$ mass bound from the most constraining 8 and 13 TeV searches we could find. The union of the exclusions from all of the searches is shown in the last line.}
\label{fig:SearchSummaries}
\end{figure}

The results of our recasting are summarized in 
Fig.~\ref{fig:SearchSummaries}.  This is the main result of our
paper.  
%
The top line of each plot (colored in blue) shows the constraints on the 
model coming from searches for resonant dilepton production. As discussed 
in the previous section, this depends only on if the $\drho$ can decay to 
leptons or not, and is independent of how the $\dpi$ decay. 
The $x$ axis for the plots is $m_{\dpi}$, so the results are obtained from 
Fig.~\ref{fig:Dileptonlimit}  by scaling the 
$x$ axis by the ratio $m_{\dpi}/m_{\drho}$.

The next two lines in the Fig.~\ref{fig:SearchSummaries} display the best 
constraints we could find for 13 TeV searches. The first of these is a 
search for supersymmetry in final states with either same-sign leptons or 
three leptons. Recasted in terms of dark pions, it excludes $m_{\dpi}$ in the $200$-$300$~GeV range for the 
gaugephilic and slightly worse for the gaugephobic categories when $\eta 
= 0.45$. This search does not work when $\eta = 0.25$ because for 
fixed $m_{\dpi}$, smaller $\eta$ implies a heavier $\rho$ and therefore a 
smaller resonant contribution to pion pair production. The other 13 TeV search 
with moderate sensitivity is a supersymmetry search with final states of 
tau leptons. The bounds from this search limit the dark pion mass in all models with $\eta < 0.5$ that we examined to be $\gtrsim 130\,\gev$, the mass above which $\dpi^+ \rightarrow \tau^+ \nu$ ceases to be the dominant decay mode.

The remaining lines in Fig.~\ref{fig:SearchSummaries} come from the 8 TeV searches which have sensitivity to $\pi_D$.
Two are multilepton searches from ATLAS and CMS, which are 
general searches counting the numbers of events for many signal regions. 
These work well for the models at low masses, and are slightly better for 
the gaugephilic models. The other exclusion comes from a search for 
supersymmetry in states with same sign leptons. In particular, one of the 
signal regions trades the usual missing energy requirement for more 
b-jets, which works well for the gaugephobic models.

Finally, the last line (shown in red) combines all of the previous 
constraints in the most naive method. The models where the $\drho$ {\em cannot} 
decay to $\dpi$ are excluded to over $m_{\dpi} = 1100\gev$ for $SU(2)$ 
kinetic mixing and to 900 GeV for $U(1)$ ($SU(2)_R$ model).  If the mass ratio allows for 
decays to pions, the exclusion limits are drastically reduced. For 
$m_{\dpi}/m_{\drho}=0.45$, the gaugephilic limits are to around 425 GeV 
while the gaugephobic limits are at 500 GeV for $SU(2)$ mixing. This 
corresponds to 13 TeV cross sections of 600 fb and 300 fb, respectively. 
It is surprising that processes with such distinct final states are still 
allowed with these large of rates at the LHC\@. The $SU(2)_R$ model limits are $m_{\pi_D} \gtrsim 130$~GeV, with a cross section of a few pb. As the mass ratio is 
further extended, the decay products become more energetic, boosting some 
of the search efficiencies. However, the resulting decrease in the cross 
section from the heavier $\drho$ compensates for this and leads to reduced limits. All of the 
models with $m_{\dpi}/m_{\drho} = 0.25$ have limits at or below 
$m_{\dpi}=200\gev$, corresponding to a (13 TeV) cross section of around a pb.
 
The rates that are still allowed are much larger than one would expect, 
especially given the exotic combinations of final state particles. In the 
next subsections, we examine the constraining searches in more detail,
looking at why the searches work and what the deficiencies are. The details we expose, combined
with the information in Sec.~\ref{sec:Additional}, will help us identify important elements that future searches should incorporate in order to improve sensitivity to dark pion scenarios.

\subsection{Searching for taus}
\label{sec:DiTau}

Working from the bottom up of the dark pion mass range, $\mathcal O(100-150\,\gev)$ dark pions in all of our benchmark models decay primarily as $\dpi^+ \rightarrow \tau^+ \nu_{\tau}$. Therefore, we begin our survey of experimental searches with searches that explicitly
look for taus.

ATLAS searches for supersymmetry in electroweak production of 
supersymmetric particles with final states with $\tau$ leptons using 14.8 
fb$^{-1}$ of $\sqrt{s}=13\tev$ data~\cite{ATLAS-CONF-2016-093}. They 
interpret the search in terms of the leptons coming from the decays of 
charginos or neutralinos. As this search is aimed at a supersymmetric 
model with a neutralino also in the final state, they require a large 
amount of missing energy, which limits the sensitivity to our benchmarks. The 
general search strategy is:
\begin{enumerate}
\item Trigger on events with two hadronically decaying $\tau$s with $p_T > 
35(25)$ GeV and have $E_T^{\text{miss}} > 50$~GeV\@.
\item Require opposite sign taus with $m_{\tau\tau} > 12$~GeV\@.
\item Veto any event with a b-jet to suppress top-quark backgrounds.
\item Suppress SM backgrounds with a $Z$ boson by removing events 
with $|m_{\tau\tau} - 79\gev| < 10\gev$.\footnote{$79$~GeV is the 
``visible'' mass of the Z for tau decays which have inherent 
missing energy.}
\item Large missing energy cut, $E_T^{\text{miss}} > 150$~GeV\@.
\item Large \emph{stransverse} mass $m_{T2} > 70$~GeV\@.
\end{enumerate}
The stransverse mass is defined as
\begin{equation}
m_{T2} = \min_{\mathbf{q_T}} \left[ \max \left(m_{T,\tau 1} (\mathbf{p}_{T,\tau1},\mathbf{q}_T ),  m_{T,\tau 2} (\mathbf{p}_{T,\tau2},\mathbf{p}_T^{miss} - \mathbf{q}_T ) \right) \right],
\end{equation}
where the transverse momenta of the two taus are $\mathbf{p}_{T,\tau1(2)}$ 
and $\mathbf{q}_T$ is the transverse vector which minimizes the larger of 
the two transverse masses. The transverse mass is defined as,
\begin{equation}
m_T (\mathbf{p}_T, \mathbf{q}_T) = \sqrt{2(p_T q_T - \mathbf{p}_T \cdot \mathbf{q}_T )} ~.
\end{equation}

\begin{figure}[t]
\begin{center}
\includegraphics[width=0.9\columnwidth]{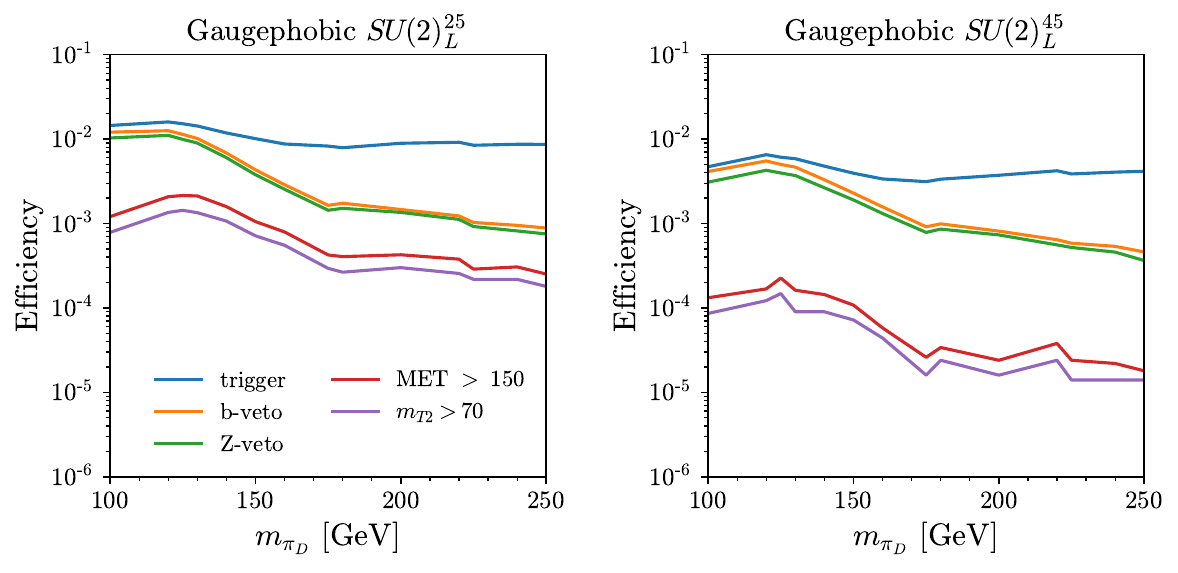}
\end{center}
\caption{Cut flow for the search for hadronically decaying taus, 
optimized for electroweak production of supersymmetric particles
\cite{ATLAS-CONF-2016-093}. The efficiency is much larger for the $\eta=0.25$
benchmarks than the $\eta=0.45$ models because the larger $\drho$ mass leads 
to more energetic $\dpi$. This increase in efficiency is offset by the decrease
in resonant production cross section.}
\label{fig:DiTauCutflow}
\end{figure}

Figure~\ref{fig:DiTauCutflow} shows the efficiency of the signal as the various cuts are being made,
and exemplifies the kinematic differences between models with different value of $\eta$.
There is very little loss in efficiency from the b- and 
$Z$-vetos for masses less than 150 GeV\@. 
Additionally, the figure shows that at this stage, there is very little difference between the $\eta$ values.
However, there is a huge drop in efficiency when requiring 
large amounts of missing energy. This is not as dramatic in the 
$\eta=0.25$ models, which produce more energetic $\dpi$ because of the heavier $\drho$. 

\begin{figure}[t]
\begin{center}
\includegraphics[width=0.7\columnwidth]{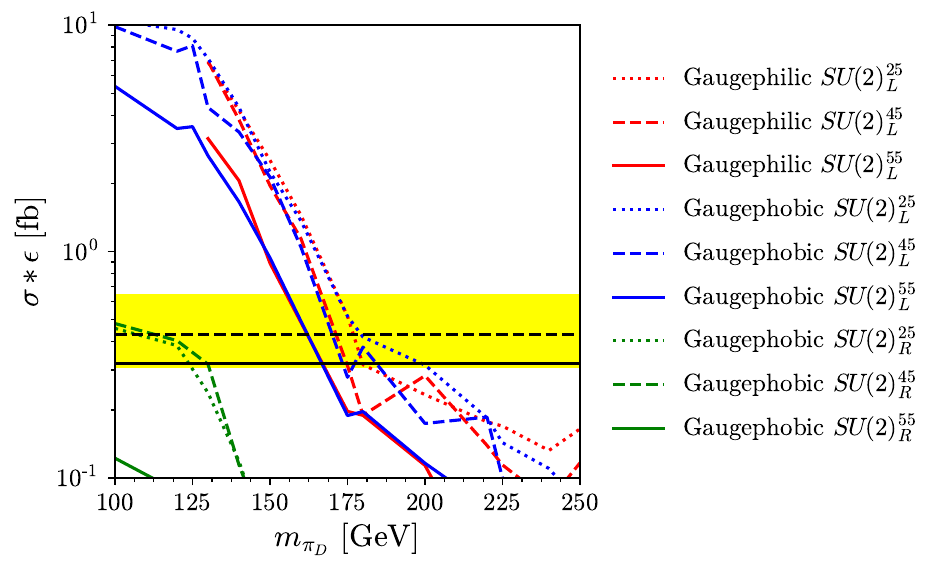}
\end{center}
\caption{Exclusions from the ATLAS search for supersymmetry in final states
with tau leptons~\cite{ATLAS-CONF-2016-093}. The dark mesons on the lighter
side of our spectrum predominately decay to taus, and the cross sections are
large. The $SU(2)_L$ type models are excluded if $m_{\dpi} \lesssim 180\gev$ while the 
$SU(2)_R$ models limits are around 130 GeV\@. This is the only search which 
limits the $SU(2)_R$ models where the $\drho$ can decay to $\dpi\dpi$.}
\label{fig:DiTauExclusions}
\end{figure}

The exclusions from this search are plotted in 
Fig.~\ref{fig:DiTauExclusions}, where the $y$-axis is
the cross section times search efficiency. The expected number of events in the 
signal region from standard model backgrounds was $5.9\pm 2.1$, while only 
three events were actually observed. As fewer events were seen than 
expected, the observed limits of 0.32~fb is more stringent than the expected 
$0.43^{+0.21}_{-0.12}$ fb. Both the gaugephilic and gaugephobic models 
with $SU(2)$ kinetic mixing and $\eta=0.25$ or $\eta=0.45$ are excluded 
from this search if $m_{\dpi} \lesssim 170-180 \gev$. Surprisingly, this search also 
constrains the $SU(2)$ models with $\eta=0.55$ even though the $\dpi$ are not produced through a resonant $\drho$.
These are only allowed if $m_{\dpi}>160\gev$.\footnote{While all of our signal numbers were determined using \texttt{Delphes} tagging and identification efficiencies, we derive limits by comparing them with ATLAS/CMS background numbers computed with their own dedicated programs and setting. As the identification and tagging efficiencies in \texttt{Delphes} are only an approximation to the true ATLAS/CMS numbers, our signal vs. background comparison is not totally genuine. To quantify the effect of the mismatch, we have checked the ramifications of changing the \texttt{Delphes} lepton identification efficiency by $\pm 10\%$ and find that this variation only leads to very minor shifts in the derived $m_{\pi_D}$ limits.}

Additionally, the $SU(2)_R$ models [that kinetically mix through $U(1)_Y$] 
with $\eta=0.25,0.45$ 
are also constrained to be above $m_{\dpi} \gtrsim 130 \gev$.  As shown in the 
summary plot of Fig.~\ref{fig:SearchSummaries}, this is the only search 
we examined which had sensitivity to the $SU(2)_R^{25,45}$ models.

The reason these limits are not stronger is because the branching ratio 
to taus is decreasing rapidly as the mass of the pions increases. This is 
compensated by an increase in the expected number of $W$s, 
$Z$s, and $b$s. The next sections examine searches which
 exploit these particles.

\subsection{Generic multilepton searches}
\label{sec:multilepton}

Examining Table~\ref{tab:Regions}, once $m_{\dpi} \gtrsim 150 \gev$,
pair produced dark pions decay to lots of bottom and top quarks, along with $Z$ and $W$. It should
be expected that searches utilizing $b$s and leptons could place strong
constraints on the benchmark models. While we studied many model driven
searches and found no limits (see Sec.~\ref{sec:Additional}), model-independent 
searches proved useful. Both ATLAS and CMS have a generic search at 8 TeV based on final states with multiple leptons. (Neither collaboration has repeated the analysis at 13 TeV).

The inclusive ATLAS search looks for $3^+$ leptons \cite{Aad:2014hja}. The basic search requirements are: 1 electron or muon for 
triggering purposes ($p_T > 26\, \GeV, |\eta| < 2.5$), a second electron 
or muon with slightly looser requirements, and a third $e/\mu$ or hadronic 
$\tau$.
The events are broken into further sub-categories according to several 
kinematic variables, such as the $b$-jet multiplicity, or whether or not 
the event contains a same-flavor-opposite-sign lepton anti-lepton pair. 
The signal regions are not orthogonal, and they set bounds on the BSM 
cross section of roughly a few fb.

Applied to $\dpi$ production, we find the most constraining signal 
regions are those containing a hadronic $\tau$ and that contain $\ge 1$ 
b-jet or have low $H_{T,L}$, defined as the scalar sum of the $p_T$ of the 
three leading leptons (or $\tau$) in the event. The limits depends strongly on
the lepton and tau identification. In 
particular, the ATLAS study used only single-prong hadronic 
taus\footnote{Also, there was no dedicated $\tau$ trigger in place for 
this analysis.} in the analysis and a benchmark identification efficiency 
of $0.5$. Compared to more recent $\tau$ reconstruction 
numbers~\cite{ATLAS-CONF-2017-029} (which are in the default \texttt{Delphes} card), the ATLAS values are worse by a factor of $\sim 2$. 
We artificially imposed the reduced tau reconstruction numbers for consistency.

\begin{figure}[t]
\begin{center}
\includegraphics[width=0.85\columnwidth]{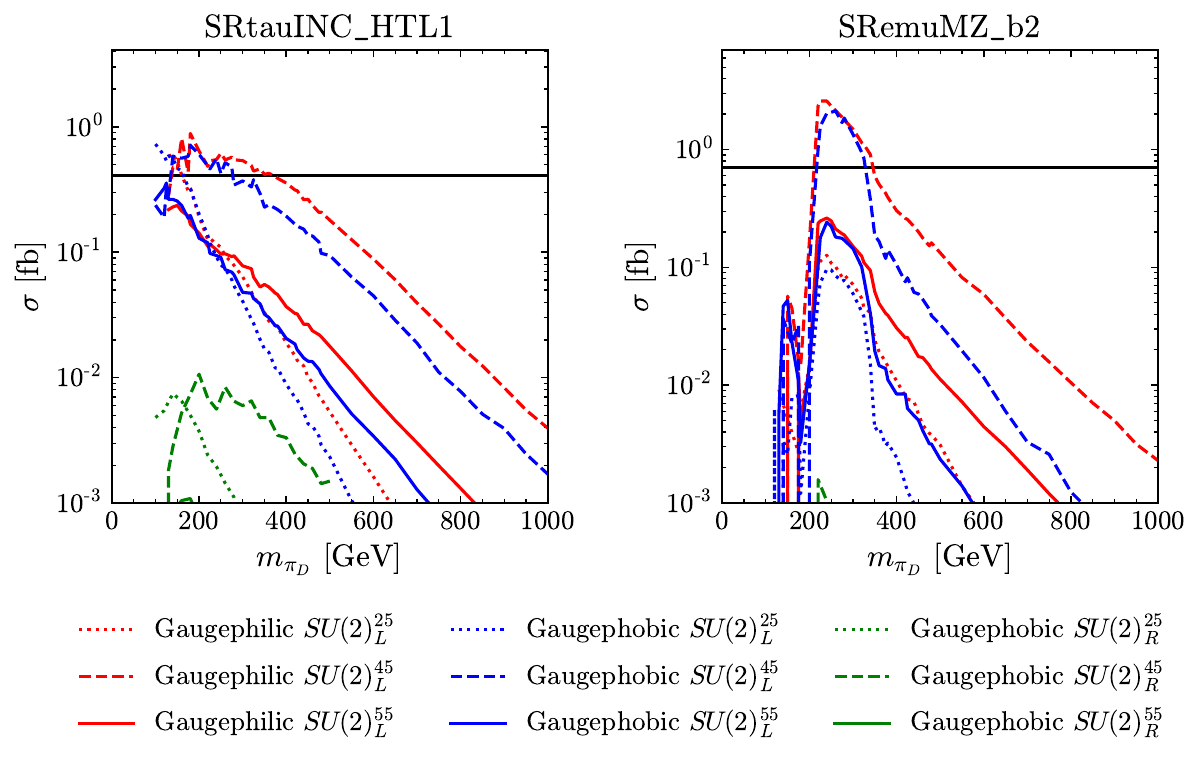}
\end{center}
\caption{Expected signal cross section in two different signal regions of the ATLAS multilepton search~\cite{Aad:2014hja} as 
a function of dark pion mass.}
\label{fig:MultiLeptonSignalRegions}
\end{figure}
The shape of the exclusion curves for two of the signal regions are shown
in Fig.~\ref{fig:MultiLeptonSignalRegions}, and exemplify the difference between
gaugephilic and gaugephobic models which were not observed in the ditau search discussed in Sec.~\ref{sec:DiTau}. 
The shapes show that the exclusions closely follow the $\pi_D$ the branching ratios.

\begin{figure}[t!]
\begin{center}
\includegraphics[width=\columnwidth]{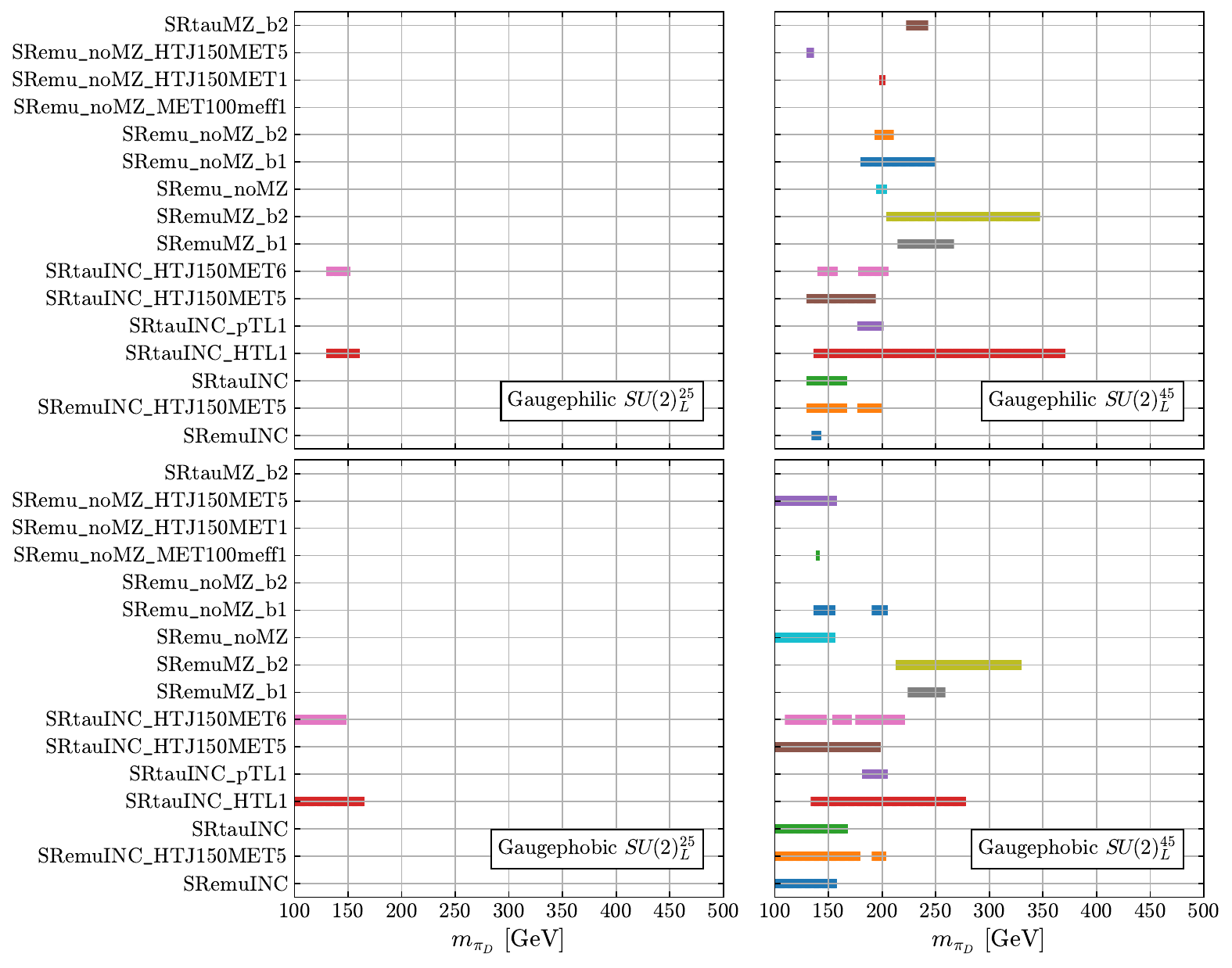}
\end{center}
\caption{Out of the 144 signal regions defined in Ref.~\cite{Aad:2014hja}, 16 regions 
constrain some portion of the dark meson parameter space. The mass ranges which
are colored are excluded. The gaugephilic models have 
larger branching ratios to $Zh$ and $Wh$ than the gaugephobic models, which leads to 
greater search efficiency and larger bounds.}
\label{fig:MLSummary}
\end{figure}
Out of the 144 signal regions defined in the ATLAS search, we find that 16 provide some level of constraint. These are summarized in Fig.~\ref{fig:MLSummary}.
Picking the strongest limit from the signal regions, we find $\pi_D > 370 \gev$ 
in the gaugephilic, $m_{\pi_D}/m_{\rho_D} = 0.45$ case and $\pi_D > 330 \gev$ 
in the gaugephobic, $m_{\pi_D}/m_{\rho_D} = 0.45$ case. For 
$m_{\pi_D}/m_{\rho_D}=0.25$ the bounds are looser, due to the fact that 
smaller $m_{\pi_D}/m_{\rho_D}$ for fixed $m_{\pi_D}$ implies a heavier 
$\rho_D$, and therefore a smaller resonant contribution to the $pp \to 
\pi_D \pi_D$ cross section. The difference between 
the limits in the gaugephilic and gaugephobic can be traced to the presence 
of more Higgs bosons in the gaugephilic $\pi_D$ decays, since 
more Higgs bosons leads to more events with $\tau$s or $b$-jets.

The CMS generic multilepton search is similar, but contains some important differences. It 
is based on 19.5 fb$^{-1}$ of $\sqrt{s}=8$ TeV data 
\cite{Chatrchyan:2014aea} and looks for events with either three or four reconstructed leptons.
In this case, the definition of leptons includes electrons with $p_T > 10\gev$ and $|\eta| < 2.4$,
muons with $p_T > 20\gev$ and $|\eta| < 2.4$,
or hadronically decaying taus with $p_T > 20\gev$ and $|\eta| < 2.3$. To trigger,
events must contain either an electron or muon with at least $p_T > 20\gev$ and events
are only allowed to have one hadronic tau.

The events are divided into 192 independent bins (96 for each of the three or four lepton cases).
The bins are split based whether there are same-flavor-opposite-sign (OSSF) pairs of 
leptons, the invariant mass of existing OSSF pair, the presence of tagged 
$b$ jets, the number of hadronic $\tau$ leptons, the amount of missing 
energy, and the scalar sum of accepted jet transverse momenta. When CMS combines their
signal regions, they are able to set bounds on new physics on the order of $\sigma\times Br \lesssim$~100~fb.

\begin{figure}[t]
\begin{center}
\includegraphics[width=0.8\columnwidth]{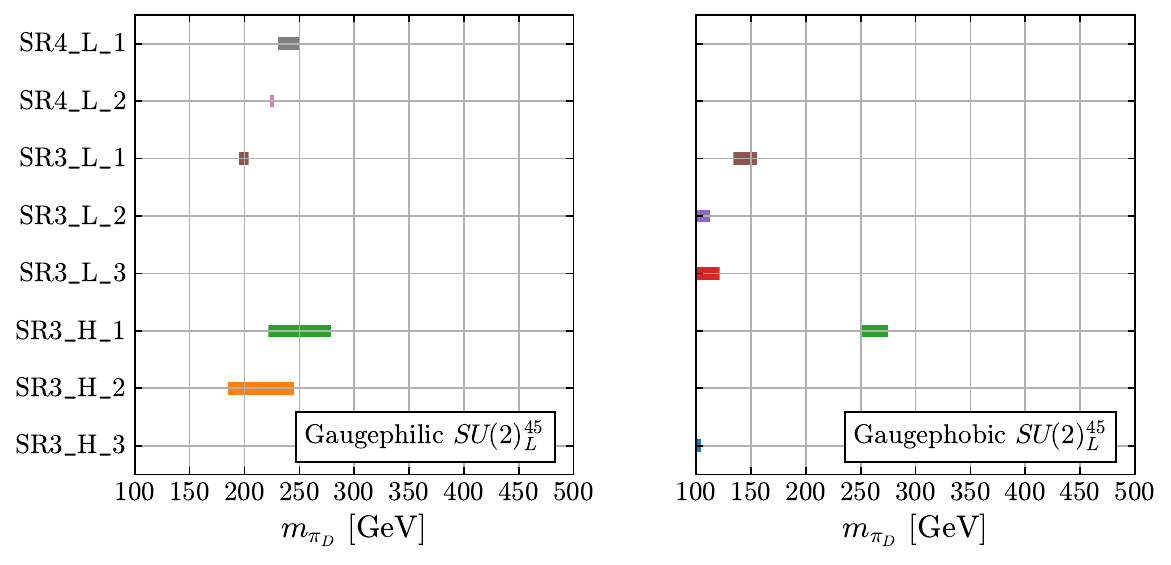}
\end{center}
\caption{Out of the 192 signal regions defined in the CMS multilepton search~\cite{Chatrchyan:2014aea}, 8 regions 
constrain some part of the dark meson parameter space. The excluded mass 
ranges are colored according to the denoted signal region. The regions labeled SR3 and SR4 regions contain either 3 or 4 leptons, respectively. The L or H denotes whether the scalar sum of the $p_T$ of the selected jets is less than 200 GeV or greater than 200 GeV. While there are different cuts concerning the number of $b$-jets or taus, all of the constraining regions require either $E_T^{\rm{miss}} < 50 \gev$ or $50\gev < E_T^{\rm{miss}} \leq 100 \gev$.}
\label{fig:CMSMLSummary}
\end{figure}

While it would in principle be possible to combine signal regions within our study, CMS does not provide the correlation information. Therefore, we are forced to examine each bin individually. This is in contrast to the method used in the ATLAS search, which used overlapping signal regions, such that some of the regions were more inclusive. Because of this, we find that the exclusions on the benchmarks from the CMS search are not as strong at the ATLAS ones. They are summarized in Fig.~\ref{fig:CMSMLSummary} for the signal regions which provide a limit. While the limits are not as strong, we find that the pattern is similar to the ATLAS result, in that the gaugephilic modes have tighter constraints than the gaugephobic models.

To date, there is no 13 TeV multi-lepton analysis. Given the success we 
see in the 8 TeV versions at catching models that fall through the cracks 
in dedicated searches (see Sec.~\ref{sec:Additional}), we encourage ATLAS and CMS to pursue similar 
model-independent, inclusive searches in the future.

\subsection{Same sign lepton searches}
\label{sec:8TeVSS}

The last type of search that we find has sensitivity to pair produced 
dark pions is also fairly generic. The main difference is that instead 
of looking for three or four leptons, they look for multiple leptons of the
same electric charge. Frustratingly, the limits we find from these scenarios
are stronger from an 8 TeV ATLAS search than the follow-up using a similar
analysis strategy at 13 TeV with more integrated luminosity.

The ATLAS search for supersymmetry using 20.7 fb$^{-1}$ of 
$\sqrt{s}=8\tev$ collisions in final states with two same sign 
leptons~\cite{ATLAS:2013tma} is a particularly powerful search. The search 
requires two leptons of the same electric charge. For electrons to be 
reconstructed, the must have $p_T > 20\gev$ and $|\eta| < 2.47$, while reconstructed muons have $p_T > 20\gev$ and $|\eta| < 2.4$. Jets are 
reconstructed with the anti-$k_t$ algorithm with a radius parameter of 0.4 
and are required to have $p_T > 40 \gev$ and $|\eta| < 2.8$. In defining 
the signal regions, the search makes use of the transverse mass, defined 
as $m_T = \sqrt{2 p_T^{\ell} E_T^{\text{miss}} \, (1-\cos\Delta \phi(\ell, 
E_T^{\text{miss}})])}$. In addition, the effective mass is defined as the 
scalar sum of the transverse momentum of the leading two leptons, the 
selected jets, and the missing energy.
 
Three different signal regions are defined. The first signal region has a 
veto on $b$-jets, which severely restricts the efficiency for higher mass 
$\dpi$. For lower masses, there is not enough missing energy in the 
events to pass the cut of $E_T^{\rm{miss}} > 150\gev$, so this signal region does not offer constraints on the model.

The next signal region looks for $\ge 1$ $b$-jet. In addition, there must 
be at least three jets (can include the $b$ jets), missing transverse 
momentum $> 150\gev$, transverse mass $>100 \gev$, and an effective mass 
$>700\gev$. There are no limits from this region as well, due to the large amount 
of missing energy required.

The third signal region takes an different approach. In addition to the 
two same sign leptons, at least three $b$ jets 
and at least 4 jets overall are required as well. In order to be statistically independent of 
the other regions, this region looks for events with small amounts of 
missing energy or transverse mass. 
The dark pions have no intrinsic 
missing energy (other than leptonic $W$ decays), but do produce a lot of 
$b$ quarks, making this an ideal signal region. 

In the gaugephobic model, the fraction of decays to $W^{\pm}\,h$ ($Z\,h$) grows with increasing charged (neutral) $\pi_D$ mass, while dark pions in the gaugephobic case predominantly decays to $t \bar{b}$ ($t\bar{t})$.  The difference in branching fractions leads to a smaller average b-jet multiplicity in the gaugephilic case which results in  a slightly lower efficiency and, as a consequence, weaker bounds. 

\begin{figure}[t]
\begin{center}
\includegraphics[width=0.9\columnwidth]{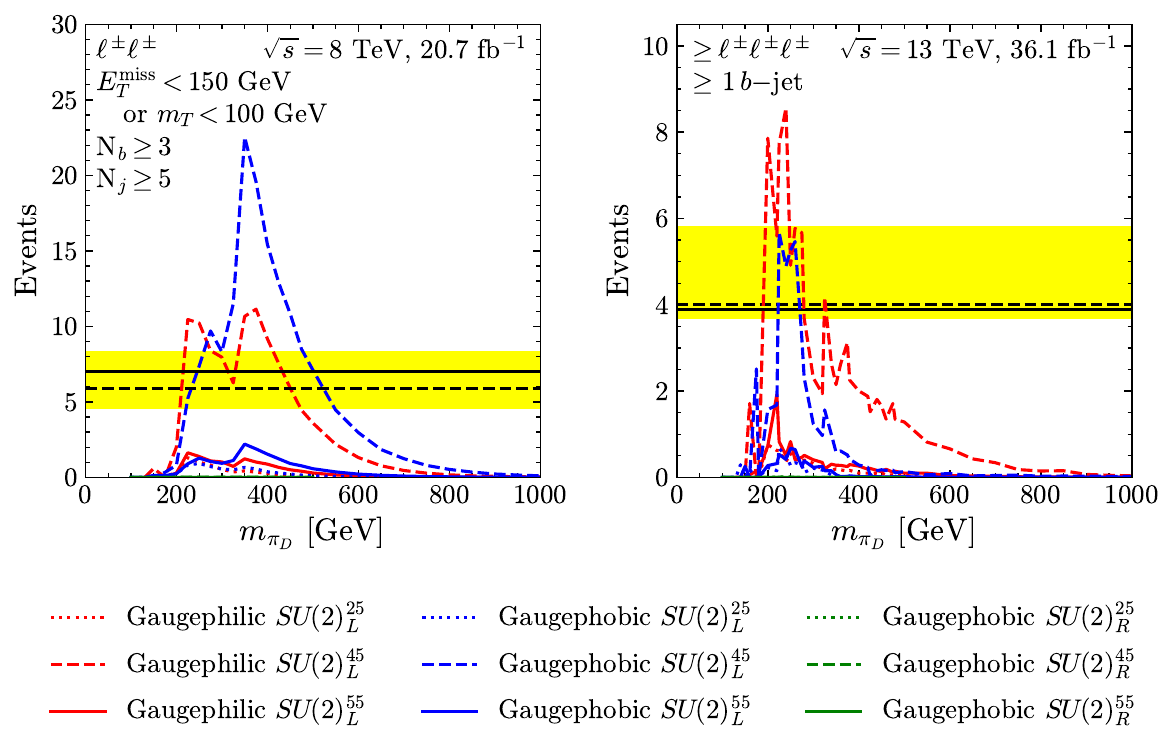}
\end{center}
\caption{Signal regions of ATLAS searches for three leptons or same sign leptons
which have sensitivity to our benchmarks. The left panel shows the limits from the 8 TeV
analysis~\cite{ATLAS:2013tma} and the right panel has the limits for the 13 TeV analysis~\cite{Aaboud:2017dmy}.
The 8 TeV analysis has bounds to the largest values of $m_{\dpi}$ for all of the 8 and 13 TeV 
analysis which we studied. The 13 TeV search does not do as well because the focus of the
analysis shifted to search for higher mass objects.
}
\label{fig:8TeVSusyLeptonsSummary}
\end{figure}

To obtain the number of expected signal events,
we multiply the cross section and luminosity by the efficiency derived
from the analysis cuts. 
These are then compared to the limits set by ATLAS\@. In the signal region, 
4 events were observed against an expected background of 
$3.1 \pm 1.6$. With this, models which would produce 7.0 expected signal events
are excluded at the 95\% CL. 
The left panel of Fig.~\ref{fig:8TeVSusyLeptonsSummary} shows the 
results of this signal region with number of expected events 
for the different models are shown in the red, blue, and green lines. The 
regions where the expected events extends above the black line are 
excluded. The only benchmarks which are limited by this search are the $SU(2)_L^{45}$ models.
The gaugephilic version is excluded for $210 \gev \lesssim m_{\dpi }\lesssim 420\gev$, while
the gaugephobic model is ruled out if $m_{\dpi}$ is between 
$250$~GeV and $500$~GeV\@. 
These are the strongest limits that we obtained for all of the searches.

With the success of the 8 TeV analysis, there was hope that when the 
search was extended to 13 TeV, the limits would greatly improve. However,
this is not that case.
Using 36 fb$^{-1}$ of $\sqrt{s}=13\tev$ collisions, ATLAS searched for 
supersymmetry in final states with two same-sign leptons or three 
leptons~\cite{Aaboud:2017dmy}. The basic requirements are nearly the
same for the lepton reconstruction, however, the $\eta$ cut is tightened to
$|\eta| < 2.0$ for electrons and loosened to $|\eta| < 2.5$ for muons.

The signal regions are more complicated in the 13 TeV analysis. There are 
19 non-exclusive signal regions 
defined in terms of the number of leptons required; the number of $b$-jets; the 
number of jets harder than 25, 40 or 50 GeV, regardless of flavor; the 
missing energy and effective mass, and the charge of the leptons.

Unlike the previous search at 8 TeV, the 13 TeV search does not have any signal 
regions which require at least three $b$-jets. Instead, to cut down on 
background, the signal regions either require more than 6 jets or large 
effective mass. This combination is aimed at TeV scale colored particles 
and does not bode well for searching for pair produced particles with 
masses in the hundreds of GeV\@.

The only one of the 19 regions that has sensitivity to heavy dark mesons is 
the one that does not have requirements on the number of jets, the 
effective mass, or the missing energy. Instead, it requires at least three 
leptons of the same-sign and one $b$-jet. In addition, it requires that no 
combination of same-sign leptons has an invariant mass around the $Z$ pole 
(veto $81 < m_{e^{\pm}e^{\pm}} < 101\gev$).

The limits from this region for the different models are shown in the right panel of
Fig.~\ref{fig:8TeVSusyLeptonsSummary}. The efficiency is largest in the mass region 
where $\dpi^{+} \rightarrow t \bar{b}$ and 
$\dpi^0 \rightarrow Zh$ dominate. The $\dpi^0 \rightarrow Zh$ mode is suppressed in the 
gaugephobic models, hence the limits are not quite as strong as the gaugephilic case. From Fig.~\ref{fig:8TeVSusyLeptonsSummary}, we see that this search only excludes $m_{\pi_D} \sim 200$-$400\,\gev$  for $\eta = 0.45$, while $\eta=0.25$ models are not constrained at all. Thus, while we 
expected that updating the best 8 TeV search would yield impressive 
bounds, it was unable to extend the limits above the 500 GeV bound set at 
$\sqrt{s}=8\tev$.

This result highlights a troubling trend.
We found the strongest limits from the 8 TeV analysis, pushing the mass 
of the dark pion to $500$~GeV for the most excluded model. 
However, that search was designed
with supersymmetry in mind, and using a supersymmetric interpretation 
of the 8 TeV search excluded sparticles (stops, specifically) up to 1 TeV\@. In the supersymmetry interpretation, it makes sense to harden the cuts and focus on particles heavier than 1 TeV\@. 
As we have seen in this analysis, however, imposing harder cuts as done with the $13$~TeV analyses is detrimental to the signals in our benchmark models, with the result that the older, $8$~TeV analyses yielded the strongest constraints.
In the next subsection, we discuss other searches which have been thwarted in a similar way.

\subsection{Additional searches}
\label{sec:Additional}

According to Table~\ref{tab:Regions} (or the branching ratios in Figs.~\ref{fig:ChargedPionBranchingRatio} and \ref{fig:NeutralPionBranchingRatio}), we expect pair produced dark pions to result in lots of third generation fermions or gauge/Higgs bosons. However, this is not a unique feature of heavy dark mesons. Many BSM scenarios involve new particles that couple predominantly to 
gauge/Higgs bosons and third generation fermions, and as a result there 
are numerous LHC searches (underway, or already done) looking for characteristics signals 
of, e.g. multiple b-jets, multiple $\tau$s, multiple $e/\mu$ in 
association with $b$-jets or $\tau$, etc. of this type of final states.
Based on energy and luminosity alone, the expectation 
is that one of these 13 TeV searches should be the most constraining. 
Our results strikingly show this is \emph{not} the case --
we find only a few searches constrain dark pions, 
with the strongest searches coming from 8 TeV\@.

Our main result, Fig.~\ref{fig:SearchSummaries} came from considering a 
wide array of BSM searches. While the details of the most successful five 
searches have been provided in the previous sections, we summarize the other, 
un-constraining searches in Table.~\ref{tab:searches}. In addition to 
the search channel, we provide a short explanation of why dark pions were 
so inefficiently captured by the search strategy.

\begin{table}[t!]
\centering
\renewcommand{\arraystretch}{1.4}
\setlength{\tabcolsep}{6pt}
\setlength{\arrayrulewidth}{.3mm}
\begin{tabular}{p{5cm}  c| p{6cm} }
\hline
\hline
Search & $\sqrt{s}$ [TeV] & Comments \\
\hline
ATLAS search for a CP-odd Higgs boson decaying to Zh  \cite{Aad:2015wra} & 8 & Veto events with more than 2 b-tagged jets kills efficiency\\ 
ATLAS search for $ t\overline{t} $ resonances \cite{Aad:2015fna} & 8 & Must have exactly one lepton. We have too many jets, confuses search\\
CMS Pair produced leptoquark \cite{Khachatryan:2014ura} & 8 & Looking for $b\bar{b} \tau^+ \tau^-$. Has minor sensitivity to overall rates, would do better with shape analysis but not enough data is provided to recast this. \\
ATLAS search for SUSY in final states with multiple b-jets \cite{Aaboud:2017hrg} & 13 & Looking for heavy states, so demands large $E_{T}^{\rm{miss}}$ and $m_{eff}$\\
CMS search for $Vh$ \cite{Khachatryan:2016cfx} & 13 & Looking for single production. Needs very boosted hard object.\\
CMS Di-Higgs $\rightarrow \tau\tau \, b\bar{b}$ \cite{Sirunyan:2017djm} & 13 & Neutral pions decay through mixing with the Higgs. Measurement uses BDTs and is not recastable. \\
CMS Low mass vector resonances $\rightarrow q\bar{q}$ \cite{Sirunyan:2017nvi} & 13& Looks for a bump on the falling soft-drop jet mass spectrum. Not enough information to recast the designed decorrelated tagger. Only sensitive to $\sigma \gtrsim 10^{3}$ pb.\\
CMS Vector-like $T \rightarrow t\,h$ \cite{Khachatryan:2016vph} & 13 & Looking for $t\,h$ resonance, only very heavy and needs QCD production. \\
\hline
\hline
\end{tabular}
\vspace{0.5cm}
\caption{Possible search strategies which seem like they should set bounds, but have limited-to-no sensitivity. 
}
\label{tab:searches}
\end{table}

While there are varying reasons the searches in Table~\ref{tab:searches} are 
not sensitive, there are a few common themes:
\begin{enumerate}
	\item Searches expect single production. This is especially true for scalars which decay to the Higgs and gauge bosons. To cut down backgrounds, events are vetoed if there are too many objects to be only $V\,h$.
	\item Searches assume large $E_T^{\rm{miss}}$. The searches which allow for pair production assume that pair production comes from a sector preserving a $Z_2$ symmetry and that therefore result in an invisible/dark matter particle at the end of the decay chain. While dark pions in the parameter space we are interested are predominantly pair-produced,  they only decay back to SM particles.
	\item Searches at 13 
TeV have their sights set on heavier new physics.  As a result, their cuts are too high to 
capture lighter dark pions. Heavier dark pions do have higher 
efficiency, but are not produced at rates the ATLAS and CMS are sensitive to, especially given that there are no leading order QCD-mediated production modes.
	\item Data is not presented in a way that is recast-friendly. For instance, the CMS pair produced leptoquark search actually has some minor sensitivity when only using the total number of events. The search then uses the shape of the scalar sum of the $p_T$ of the light lepton, the hadronic $\tau$, and the two jets to set limits, but they do not provide a fit of the shape. Similarly, experiments trying to measure standard model processes (such as $hh\rightarrow b\bar{b} \tau^+\tau^-$) may potentially be sensitive to some $\pi_D$ parameter space, but  they rely on machine learning techniques which cannot be reproduced.
\end{enumerate}

We encourage the experiments to continue to push the limits of the LHC searches using all of the techniques they have available. However, as it is not possible for them to test every theory model, it is important that the results be presented in such a way that they can be reproduced without insider knowledge.

\section{Conclusions}
\label{sec:conclusions}

\begin{itemize}
\item In this paper we have examined the phenomenology of dark pions -- composite states with electroweak and Higgs interactions that may lurk at the electroweak scale. Dark pion - like states are a component of many BSM scenarios with new strong dynamics near the electroweak scale. 
\item In addition to electroweak interactions, dark pions are also resonantly produced via dark rhos that kinetically mix with SM gauge bosons and decay through interactions with SM fermions or into $hV$. The overall size of the single-pion to SM coupling and the relative strength of the fermionic versus $Vh$ decay modes encodes some information about the symmetry structure of the strong sector and is the subject of Ref.~\cite{Kribs:2018oad}.
\item Taken more abstractly, dark pions represent a type of new physics that is predominantly pair produced, is uncolored, and decays back to SM final states. This is a particularly tricky combination for the LHC, since the lack of strong interactions means the BSM cross sections are small and the fact that the final states are pure SM leaves few easy handles to separate signal from background.
\item The phenomenology of the dark pions is governed largely by a few parameters; the relative strength of the dark pion decays to  fermionic versus gauge bosons, the type of kinetic mixing [whether with $SU(2)_L$ or $U(1)_Y$], and the mass of $\pi_D$ relative to $\rho_D$.  Setting up nine benchmark models with different values for these key parameters, we explored the constraints on dark mesons from 8 and 13 TeV LHC searches.
\item The only scenario where we find constraints in the TeV range is when the $\rho^0_D$ is kinematically forbidden from decaying to dark pions and therefore decays with significant branching ratio into leptons, the $SU(2)^{55}_{L,R}$ cases. 
For all other cases, $\rho_D \to \pi_D \pi_D$ is kinematically accessible so the dilepton bounds are negligible and the best avenue is to look for signals of $\pi_D$ pairs. Depending on the type of kinetic mixing and the relative mass of the $\rho_D$ mesons, the bounds on $m_{\pi_D}$ from $\pi_D$ pair production signals vary from slightly above the LEP II charged particle bound to $\sim 500\, \gev$. The strongest bounds come when the mass of $\rho_D$ is not too much heavier than $2\, m_{\pi_D}$, and kinetically mix with the $SU(2)_L$, while the weakest bounds come when the kinetic mixing only involves $U(1)_Y$. As the most extreme example of how light these particles can be while remaining undetected, consider the $SU(2)^{45}_{R}$ model. There, dark pions as light as $\sim 130\, \gev$ are still viable; perhaps more surprising, the vector $\rho_D$ in this scenario sits at $\sim 300\, \gev$!
\item In our survey of LHC searches, we found the most useful features for bounding dark mesons to be signal regions with high multiplicity of leptons and/or b-jets  {\em without} strong requirements on the energy (of the individual objects, or summed) or missing energy. As model-specific searches march towards higher masses in the 13 TeV era, this type of signal region has become rarer and rarer. For scenarios without a dedicated search, such as the dark meson explored here  -- or, more generally, for types of BSM physics that is pair produced with sub-QCD rates and does not bring a non-SM source of missing energy -- the net result is that 13 TeV searches can be less sensitive than 8 TeV versions.  Generic searches based on multiple leptons served as a catch-all for this type of ``non-standard'' BSM at 8 TeV, and we encourage ATLAS and CMS to repeat similar studies with 13 TeV\@.

\end{itemize}

\section*{Acknowledgments}

We thank S.~Chang and E.~Neil for helpful 
discussions as this paper was being completed.  
The work utilized the University of Oregon Talapas high performance  
computing cluster.  
GDK and AM thank the Universities Research Association for 
travel support, and Fermilab for hospitality, where part of 
this work was completed.
This work of GDK, BO and TT was supported in part by the 
U.S. Department of Energy under Grant Number DE-SC0011640.
The work of AM was supported in part by  
the National Science Foundation  
under Grant Numbers PHY-1520966 and PHY-1820860. 

\bibliographystyle{utphys}
\bibliography{DP}

\providecommand{\href}[2]{#2}\begingroup\raggedright\begin{thebibliography}{100}

\bibitem{Simmons:1988fu}
E.~H. Simmons, ``{Phenomenology of a Technicolor Model With Heavy Scalar
  Doublet},''
\href{http://dx.doi.org/10.1016/0550-3213(89)90296-4}{{\em Nucl. Phys.} {\bf
  B312} (1989)  253--268}.

\bibitem{Samuel:1990dq}
S.~Samuel, ``{BOSONIC TECHNICOLOR},''
\href{http://dx.doi.org/10.1016/0550-3213(90)90378-Q}{{\em Nucl. Phys.} {\bf
  B347} (1990)  625--650}.

\bibitem{Dine:1990jd}
M.~Dine, A.~Kagan, and S.~Samuel, ``{Naturalness in Supersymmetry, or Raising
  the Supersymmetry Breaking Scale},''
\href{http://dx.doi.org/10.1016/0370-2693(90)90847-Y}{{\em Phys. Lett.} {\bf
  B243} (1990)  250--256}.

\bibitem{Kagan:1990az}
A.~Kagan and S.~Samuel, ``{The Family mass hierarchy problem in bosonic
  technicolor},''
\href{http://dx.doi.org/10.1016/0370-2693(90)90492-O}{{\em Phys. Lett.} {\bf
  B252} (1990)  605--610}.

\bibitem{Kagan:1991gh}
A.~Kagan and S.~Samuel, ``{Renormalization group aspects of bosonic
  technicolor},''
\href{http://dx.doi.org/10.1016/0370-2693(91)91535-4}{{\em Phys. Lett.} {\bf
  B270} (1991)  37--44}.

\bibitem{Carone:1992rh}
C.~D. Carone and E.~H. Simmons, ``{Oblique corrections in technicolor with a
  scalar},'' \href{http://dx.doi.org/10.1016/0550-3213(93)90187-T}{{\em Nucl.
  Phys.} {\bf B397} (1993)  591--615},
\href{http://arxiv.org/abs/hep-ph/9207273}{{\tt arXiv:hep-ph/9207273
  [hep-ph]}}.

\bibitem{Carone:1993xc}
C.~D. Carone and H.~Georgi, ``{Technicolor with a massless scalar doublet},''
  \href{http://dx.doi.org/10.1103/PhysRevD.49.1427}{{\em Phys. Rev.} {\bf D49}
  (1994)  1427--1436},
\href{http://arxiv.org/abs/hep-ph/9308205}{{\tt arXiv:hep-ph/9308205
  [hep-ph]}}.

\bibitem{Dobrescu:1997kt}
B.~A. Dobrescu and J.~Terning, ``{Negative contributions to S in an effective
  field theory},'' \href{http://dx.doi.org/10.1016/S0370-2693(97)01304-X}{{\em
  Phys. Lett.} {\bf B416} (1998)  129--136},
\href{http://arxiv.org/abs/hep-ph/9709297}{{\tt arXiv:hep-ph/9709297
  [hep-ph]}}.

\bibitem{Antola:2009wq}
M.~Antola, M.~Heikinheimo, F.~Sannino, and K.~Tuominen, ``{Unnatural Origin of
  Fermion Masses for Technicolor},''
  \href{http://dx.doi.org/10.1007/JHEP03(2010)050}{{\em JHEP} {\bf 03} (2010)
  050},
\href{http://arxiv.org/abs/0910.3681}{{\tt arXiv:0910.3681 [hep-ph]}}.

\bibitem{Azatov:2011ht}
A.~Azatov, J.~Galloway, and M.~A. Luty, ``{Superconformal Technicolor},''
  \href{http://dx.doi.org/10.1103/PhysRevLett.108.041802}{{\em Phys. Rev.
  Lett.} {\bf 108} (2012)  041802},
\href{http://arxiv.org/abs/1106.3346}{{\tt arXiv:1106.3346 [hep-ph]}}.

\bibitem{Azatov:2011ps}
A.~Azatov, J.~Galloway, and M.~A. Luty, ``{Superconformal Technicolor: Models
  and Phenomenology},''
  \href{http://dx.doi.org/10.1103/PhysRevD.85.015018}{{\em Phys. Rev.} {\bf
  D85} (2012)  015018},
\href{http://arxiv.org/abs/1106.4815}{{\tt arXiv:1106.4815 [hep-ph]}}.

\bibitem{Gherghetta:2011na}
T.~Gherghetta and A.~Pomarol, ``{A Distorted MSSM Higgs Sector from Low-Scale
  Strong Dynamics},'' \href{http://dx.doi.org/10.1007/JHEP12(2011)069}{{\em
  JHEP} {\bf 12} (2011)  069},
\href{http://arxiv.org/abs/1107.4697}{{\tt arXiv:1107.4697 [hep-ph]}}.

\bibitem{Galloway:2013dma}
J.~Galloway, M.~A. Luty, Y.~Tsai, and Y.~Zhao, ``{Induced Electroweak Symmetry
  Breaking and Supersymmetric Naturalness},''
  \href{http://dx.doi.org/10.1103/PhysRevD.89.075003}{{\em Phys. Rev.} {\bf
  D89} (2014) no.~7, 075003},
\href{http://arxiv.org/abs/1306.6354}{{\tt arXiv:1306.6354 [hep-ph]}}.

\bibitem{Chang:2014ida}
S.~Chang, J.~Galloway, M.~Luty, E.~Salvioni, and Y.~Tsai, ``{Phenomenology of
  Induced Electroweak Symmetry Breaking},''
  \href{http://dx.doi.org/10.1007/JHEP03(2015)017}{{\em JHEP} {\bf 03} (2015)
  017},
\href{http://arxiv.org/abs/1411.6023}{{\tt arXiv:1411.6023 [hep-ph]}}.

\bibitem{Beauchesne:2015lva}
H.~Beauchesne, K.~Earl, and T.~Gregoire, ``{The spontaneous Z2 breaking Twin
  Higgs},'' \href{http://dx.doi.org/10.1007/JHEP01(2016)130}{{\em JHEP} {\bf
  01} (2016)  130},
\href{http://arxiv.org/abs/1510.06069}{{\tt arXiv:1510.06069 [hep-ph]}}.

\bibitem{Harnik:2016koz}
R.~Harnik, K.~Howe, and J.~Kearney, ``{Tadpole-Induced Electroweak Symmetry
  Breaking and pNGB Higgs Models},''
  \href{http://dx.doi.org/10.1007/JHEP03(2017)111}{{\em JHEP} {\bf 03} (2017)
  111},
\href{http://arxiv.org/abs/1603.03772}{{\tt arXiv:1603.03772 [hep-ph]}}.

\bibitem{Alanne:2016rpe}
T.~Alanne, M.~T. Frandsen, and D.~Buarque~Franzosi, ``{Testing a dynamical
  origin of Standard Model fermion masses},''
  \href{http://dx.doi.org/10.1103/PhysRevD.94.071703}{{\em Phys. Rev.} {\bf
  D94} (2016)  071703},
\href{http://arxiv.org/abs/1607.01440}{{\tt arXiv:1607.01440 [hep-ph]}}.

\bibitem{Galloway:2016fuo}
J.~Galloway, A.~L. Kagan, and A.~Martin, ``{A UV complete partially
  composite-pNGB Higgs},''
  \href{http://dx.doi.org/10.1103/PhysRevD.95.035038}{{\em Phys. Rev.} {\bf
  D95} (2017) no.~3, 035038},
\href{http://arxiv.org/abs/1609.05883}{{\tt arXiv:1609.05883 [hep-ph]}}.

\bibitem{Agugliaro:2016clv}
A.~Agugliaro, O.~Antipin, D.~Becciolini, S.~De~Curtis, and M.~Redi, ``{UV
  complete composite Higgs models},''
  \href{http://dx.doi.org/10.1103/PhysRevD.95.035019}{{\em Phys. Rev.} {\bf
  D95} (2017) no.~3, 035019},
\href{http://arxiv.org/abs/1609.07122}{{\tt arXiv:1609.07122 [hep-ph]}}.

\bibitem{Barducci:2018yer}
D.~Barducci, S.~De~Curtis, M.~Redi, and A.~Tesi, ``{An almost elementary Higgs:
  Theory and Practice},'' \href{http://dx.doi.org/10.1007/JHEP08(2018)017}{{\em
  JHEP} {\bf 08} (2018)  017},
\href{http://arxiv.org/abs/1805.12578}{{\tt arXiv:1805.12578 [hep-ph]}}.

\bibitem{Bellazzini:2014yua}
B.~Bellazzini, C.~Csáki, and J.~Serra, ``{Composite Higgses},''
  \href{http://dx.doi.org/10.1140/epjc/s10052-014-2766-x}{{\em Eur. Phys. J.}
  {\bf C74} (2014) no.~5, 2766},
\href{http://arxiv.org/abs/1401.2457}{{\tt arXiv:1401.2457 [hep-ph]}}.

\bibitem{Graham:2015cka}
P.~W. Graham, D.~E. Kaplan, and S.~Rajendran, ``{Cosmological Relaxation of the
  Electroweak Scale},''
  \href{http://dx.doi.org/10.1103/PhysRevLett.115.221801}{{\em Phys. Rev.
  Lett.} {\bf 115} (2015) no.~22, 221801},
\href{http://arxiv.org/abs/1504.07551}{{\tt arXiv:1504.07551 [hep-ph]}}.

\bibitem{Antipin:2015jia}
O.~Antipin and M.~Redi, ``{The Half-composite Two Higgs Doublet Model and the
  Relaxion},'' \href{http://dx.doi.org/10.1007/JHEP12(2015)031}{{\em JHEP} {\bf
  12} (2015)  031},
\href{http://arxiv.org/abs/1508.01112}{{\tt arXiv:1508.01112 [hep-ph]}}.

\bibitem{Batell:2017kho}
B.~Batell, M.~A. Fedderke, and L.-T. Wang, ``{Relaxation of the Composite Higgs
  Little Hierarchy},'' \href{http://dx.doi.org/10.1007/JHEP12(2017)139}{{\em
  JHEP} {\bf 12} (2017)  139},
\href{http://arxiv.org/abs/1705.09666}{{\tt arXiv:1705.09666 [hep-ph]}}.

\bibitem{Nussinov:1985xr}
S.~Nussinov, ``{TECHNOCOSMOLOGY: COULD A TECHNIBARYON EXCESS PROVIDE A
  'NATURAL' MISSING MASS CANDIDATE?},''
\href{http://dx.doi.org/10.1016/0370-2693(85)90689-6}{{\em Phys.Lett.} {\bf
  B165} (1985)  55}.

\bibitem{Chivukula:1989qb}
R.~S. Chivukula and T.~P. Walker, ``{TECHNICOLOR COSMOLOGY},''
\href{http://dx.doi.org/10.1016/0550-3213(90)90151-3}{{\em Nucl.Phys.} {\bf
  B329} (1990)  445}.

\bibitem{Barr:1990ca}
S.~M. Barr, R.~S. Chivukula, and E.~Farhi, ``{Electroweak Fermion Number
  Violation and the Production of Stable Particles in the Early Universe},''
\href{http://dx.doi.org/10.1016/0370-2693(90)91661-T}{{\em Phys.Lett.} {\bf
  B241} (1990)  387--391}.

\bibitem{Barr:1991qn}
S.~M. Barr, ``{Baryogenesis, sphalerons and the cogeneration of dark matter},''
\href{http://dx.doi.org/10.1103/PhysRevD.44.3062}{{\em Phys.Rev.} {\bf D44}
  (1991)  3062--3066}.

\bibitem{Kaplan:1991ah}
D.~B. Kaplan, ``{A Single explanation for both the baryon and dark matter
  densities},''
\href{http://dx.doi.org/10.1103/PhysRevLett.68.741}{{\em Phys.Rev.Lett.} {\bf
  68} (1992)  741--743}.

\bibitem{Chivukula:1992pn}
R.~S. Chivukula, A.~G. Cohen, M.~E. Luke, and M.~J. Savage, ``{A Comment on the
  strong interactions of color - neutral technibaryons},''
  \href{http://dx.doi.org/10.1016/0370-2693(93)91836-C}{{\em Phys.Lett.} {\bf
  B298} (1993)  380--382},
\href{http://arxiv.org/abs/hep-ph/9210274}{{\tt arXiv:hep-ph/9210274
  [hep-ph]}}.

\bibitem{Bagnasco:1993st}
J.~Bagnasco, M.~Dine, and S.~D. Thomas, ``{Detecting technibaryon dark
  matter},'' \href{http://dx.doi.org/10.1016/0370-2693(94)90830-3}{{\em
  Phys.Lett.} {\bf B320} (1994)  99--104},
\href{http://arxiv.org/abs/hep-ph/9310290}{{\tt arXiv:hep-ph/9310290
  [hep-ph]}}.

\bibitem{Khlopov:2008ty}
M.~{\relax Yu}. Khlopov and C.~Kouvaris, ``{Composite dark matter from a model
  with composite Higgs boson},''
  \href{http://dx.doi.org/10.1103/PhysRevD.78.065040}{{\em Phys. Rev.} {\bf
  D78} (2008)  065040},
\href{http://arxiv.org/abs/0806.1191}{{\tt arXiv:0806.1191 [astro-ph]}}.

\bibitem{Ryttov:2008xe}
T.~A. Ryttov and F.~Sannino, ``{Ultra Minimal Technicolor and its Dark Matter
  TIMP},'' \href{http://dx.doi.org/10.1103/PhysRevD.78.115010}{{\em Phys. Rev.}
  {\bf D78} (2008)  115010},
\href{http://arxiv.org/abs/0809.0713}{{\tt arXiv:0809.0713 [hep-ph]}}.

\bibitem{Hambye:2009fg}
T.~Hambye and M.~H.~G. Tytgat, ``{Confined hidden vector dark matter},''
  \href{http://dx.doi.org/10.1016/j.physletb.2009.11.050}{{\em Phys. Lett.}
  {\bf B683} (2010)  39--41},
\href{http://arxiv.org/abs/0907.1007}{{\tt arXiv:0907.1007 [hep-ph]}}.

\bibitem{Bai:2010qg}
Y.~Bai and R.~J. Hill, ``{Weakly Interacting Stable Pions},''
  \href{http://dx.doi.org/10.1103/PhysRevD.82.111701}{{\em Phys.Rev.} {\bf D82}
  (2010)  111701},
\href{http://arxiv.org/abs/1005.0008}{{\tt arXiv:1005.0008 [hep-ph]}}.

\bibitem{Lewis:2011zb}
R.~Lewis, C.~Pica, and F.~Sannino, ``{Light Asymmetric Dark Matter on the
  Lattice: SU(2) Technicolor with Two Fundamental Flavors},''
  \href{http://dx.doi.org/10.1103/PhysRevD.85.014504}{{\em Phys.Rev.} {\bf D85}
  (2012)  014504},
\href{http://arxiv.org/abs/1109.3513}{{\tt arXiv:1109.3513 [hep-ph]}}.

\bibitem{Frigerio:2012uc}
M.~Frigerio, A.~Pomarol, F.~Riva, and A.~Urbano, ``{Composite Scalar Dark
  Matter},'' \href{http://dx.doi.org/10.1007/JHEP07(2012)015}{{\em JHEP} {\bf
  07} (2012)  015},
\href{http://arxiv.org/abs/1204.2808}{{\tt arXiv:1204.2808 [hep-ph]}}.

\bibitem{Buckley:2012ky}
M.~R. Buckley and E.~T. Neil, ``{Thermal Dark Matter from a Confining
  Sector},'' \href{http://dx.doi.org/10.1103/PhysRevD.87.043510}{{\em
  Phys.Rev.} {\bf D87} (2013) no.~4, 043510},
\href{http://arxiv.org/abs/1209.6054}{{\tt arXiv:1209.6054 [hep-ph]}}.

\bibitem{Bhattacharya:2013kma}
S.~Bhattacharya, B.~Melić, and J.~Wudka, ``{Pionic Dark Matter},''
  \href{http://dx.doi.org/10.1007/JHEP02(2014)115}{{\em JHEP} {\bf 02} (2014)
  115},
\href{http://arxiv.org/abs/1307.2647}{{\tt arXiv:1307.2647 [hep-ph]}}.

\bibitem{Marzocca:2014msa}
D.~Marzocca and A.~Urbano, ``{Composite Dark Matter and LHC Interplay},''
  \href{http://dx.doi.org/10.1007/JHEP07(2014)107}{{\em JHEP} {\bf 07} (2014)
  107},
\href{http://arxiv.org/abs/1404.7419}{{\tt arXiv:1404.7419 [hep-ph]}}.

\bibitem{Hietanen:2014xca}
A.~Hietanen, R.~Lewis, C.~Pica, and F.~Sannino, ``{Fundamental Composite Higgs
  Dynamics on the Lattice: SU(2) with Two Flavors},''
  \href{http://dx.doi.org/10.1007/JHEP07(2014)116}{{\em JHEP} {\bf 07} (2014)
  116},
\href{http://arxiv.org/abs/1404.2794}{{\tt arXiv:1404.2794 [hep-lat]}}.

\bibitem{Pasechnik:2014ida}
R.~Pasechnik, V.~Beylin, V.~Kuksa, and G.~Vereshkov, ``{Composite scalar Dark
  Matter from vector-like $SU(2)$ confinement},''
  \href{http://dx.doi.org/10.1142/S0217751X16500366}{{\em Int. J. Mod. Phys.}
  {\bf A31} (2016) no.~08, 1650036},
\href{http://arxiv.org/abs/1407.2392}{{\tt arXiv:1407.2392 [hep-ph]}}.

\bibitem{Antipin:2014qva}
O.~Antipin, M.~Redi, and A.~Strumia, ``{Dynamical generation of the weak and
  Dark Matter scales from strong interactions},''
  \href{http://dx.doi.org/10.1007/JHEP01(2015)157}{{\em JHEP} {\bf 01} (2015)
  157},
\href{http://arxiv.org/abs/1410.1817}{{\tt arXiv:1410.1817 [hep-ph]}}.

\bibitem{Hochberg:2014kqa}
Y.~Hochberg, E.~Kuflik, H.~Murayama, T.~Volansky, and J.~G. Wacker, ``{Model
  for Thermal Relic Dark Matter of Strongly Interacting Massive Particles},''
  \href{http://dx.doi.org/10.1103/PhysRevLett.115.021301}{{\em Phys. Rev.
  Lett.} {\bf 115} (2015) no.~2, 021301},
\href{http://arxiv.org/abs/1411.3727}{{\tt arXiv:1411.3727 [hep-ph]}}.

\bibitem{Carmona:2015haa}
A.~Carmona and M.~Chala, ``{Composite Dark Sectors},''
  \href{http://dx.doi.org/10.1007/JHEP06(2015)105}{{\em JHEP} {\bf 06} (2015)
  105},
\href{http://arxiv.org/abs/1504.00332}{{\tt arXiv:1504.00332 [hep-ph]}}.

\bibitem{Lee:2015gsa}
H.~M. Lee and M.-S. Seo, ``{Communication with SIMP dark mesons via Z'
  -portal},'' \href{http://dx.doi.org/10.1016/j.physletb.2015.07.013}{{\em
  Phys. Lett.} {\bf B748} (2015)  316--322},
\href{http://arxiv.org/abs/1504.00745}{{\tt arXiv:1504.00745 [hep-ph]}}.

\bibitem{Hochberg:2015vrg}
Y.~Hochberg, E.~Kuflik, and H.~Murayama, ``{SIMP Spectroscopy},''
  \href{http://dx.doi.org/10.1007/JHEP05(2016)090}{{\em JHEP} {\bf 05} (2016)
  090},
\href{http://arxiv.org/abs/1512.07917}{{\tt arXiv:1512.07917 [hep-ph]}}.

\bibitem{Bruggisser:2016ixa}
S.~Bruggisser, F.~Riva, and A.~Urbano, ``{Strongly Interacting Light Dark
  Matter},'' \href{http://dx.doi.org/10.21468/SciPostPhys.3.3.017}{{\em SciPost
  Phys.} {\bf 3} (2017) no.~3, 017},
\href{http://arxiv.org/abs/1607.02474}{{\tt arXiv:1607.02474 [hep-ph]}}.

\bibitem{Ma:2017vzm}
Y.~Wu, T.~Ma, B.~Zhang, and G.~Cacciapaglia, ``{Composite Dark Matter and
  Higgs},'' \href{http://dx.doi.org/10.1007/JHEP11(2017)058}{{\em JHEP} {\bf
  11} (2017)  058},
\href{http://arxiv.org/abs/1703.06903}{{\tt arXiv:1703.06903 [hep-ph]}}.

\bibitem{Davoudiasl:2017zws}
H.~Davoudiasl, P.~P. Giardino, E.~T. Neil, and E.~Rinaldi, ``{Unified Scenario
  for Composite Right-Handed Neutrinos and Dark Matter},''
  \href{http://dx.doi.org/10.1103/PhysRevD.96.115003}{{\em Phys. Rev.} {\bf
  D96} (2017) no.~11, 115003},
\href{http://arxiv.org/abs/1709.01082}{{\tt arXiv:1709.01082 [hep-ph]}}.

\bibitem{Berlin:2018tvf}
A.~Berlin, N.~Blinov, S.~Gori, P.~Schuster, and N.~Toro, ``{Cosmology and
  Accelerator Tests of Strongly Interacting Dark Matter},''
  \href{http://dx.doi.org/10.1103/PhysRevD.97.055033}{{\em Phys. Rev.} {\bf
  D97} (2018) no.~5, 055033},
\href{http://arxiv.org/abs/1801.05805}{{\tt arXiv:1801.05805 [hep-ph]}}.

\bibitem{Choi:2018iit}
S.-M. Choi, H.~M. Lee, P.~Ko, and A.~Natale, ``{Resolving phenomenological
  problems with strongly-interacting-massive-particle models with dark vector
  resonances},'' \href{http://dx.doi.org/10.1103/PhysRevD.98.015034}{{\em Phys.
  Rev.} {\bf D98} (2018) no.~1, 015034},
\href{http://arxiv.org/abs/1801.07726}{{\tt arXiv:1801.07726 [hep-ph]}}.

\bibitem{Hochberg:2018rjs}
Y.~Hochberg, E.~Kuflik, R.~Mcgehee, H.~Murayama, and K.~Schutz, ``{SIMPs
  through the axion portal},''
\href{http://arxiv.org/abs/1806.10139}{{\tt arXiv:1806.10139 [hep-ph]}}.

\bibitem{Alves:2009nf}
D.~S. Alves, S.~R. Behbahani, P.~Schuster, and J.~G. Wacker, ``{Composite
  Inelastic Dark Matter},''
  \href{http://dx.doi.org/10.1016/j.physletb.2010.08.006}{{\em Phys.Lett.} {\bf
  B692} (2010)  323--326},
\href{http://arxiv.org/abs/0903.3945}{{\tt arXiv:0903.3945 [hep-ph]}}.

\bibitem{Kribs:2009fy}
G.~D. Kribs, T.~S. Roy, J.~Terning, and K.~M. Zurek, ``{Quirky Composite Dark
  Matter},'' \href{http://dx.doi.org/10.1103/PhysRevD.81.095001}{{\em Phys.
  Rev.} {\bf D81} (2010)  095001},
\href{http://arxiv.org/abs/0909.2034}{{\tt arXiv:0909.2034 [hep-ph]}}.

\bibitem{Lisanti:2009am}
M.~Lisanti and J.~G. Wacker, ``{Parity Violation in Composite Inelastic Dark
  Matter Models},'' \href{http://dx.doi.org/10.1103/PhysRevD.82.055023}{{\em
  Phys. Rev.} {\bf D82} (2010)  055023},
\href{http://arxiv.org/abs/0911.4483}{{\tt arXiv:0911.4483 [hep-ph]}}.

\bibitem{Alves:2010dd}
D.~Spier Moreira~Alves, S.~R. Behbahani, P.~Schuster, and J.~G. Wacker, ``{The
  Cosmology of Composite Inelastic Dark Matter},''
  \href{http://dx.doi.org/10.1007/JHEP06(2010)113}{{\em JHEP} {\bf 1006} (2010)
   113},
\href{http://arxiv.org/abs/1003.4729}{{\tt arXiv:1003.4729 [hep-ph]}}.

\bibitem{Geller:2018biy}
M.~Geller, S.~Iwamoto, G.~Lee, Y.~Shadmi, and O.~Telem, ``{Dark quarkonium
  formation in the early universe},''
  \href{http://dx.doi.org/10.1007/JHEP06(2018)135}{{\em JHEP} {\bf 06} (2018)
  135},
\href{http://arxiv.org/abs/1802.07720}{{\tt arXiv:1802.07720 [hep-ph]}}.

\bibitem{Gudnason:2006yj}
S.~B. Gudnason, C.~Kouvaris, and F.~Sannino, ``{Dark Matter from new
  Technicolor Theories},''
  \href{http://dx.doi.org/10.1103/PhysRevD.74.095008}{{\em Phys. Rev.} {\bf
  D74} (2006)  095008},
\href{http://arxiv.org/abs/hep-ph/0608055}{{\tt arXiv:hep-ph/0608055
  [hep-ph]}}.

\bibitem{Dietrich:2006cm}
D.~D. Dietrich and F.~Sannino, ``{Conformal window of SU(N) gauge theories with
  fermions in higher dimensional representations},''
  \href{http://dx.doi.org/10.1103/PhysRevD.75.085018}{{\em Phys. Rev.} {\bf
  D75} (2007)  085018},
\href{http://arxiv.org/abs/hep-ph/0611341}{{\tt arXiv:hep-ph/0611341
  [hep-ph]}}.

\bibitem{Foadi:2008qv}
R.~Foadi, M.~T. Frandsen, and F.~Sannino, ``{Technicolor Dark Matter},''
  \href{http://dx.doi.org/10.1103/PhysRevD.80.037702}{{\em Phys. Rev.} {\bf
  D80} (2009)  037702},
\href{http://arxiv.org/abs/0812.3406}{{\tt arXiv:0812.3406 [hep-ph]}}.

\bibitem{Mardon:2009gw}
J.~Mardon, Y.~Nomura, and J.~Thaler, ``{Cosmic Signals from the Hidden
  Sector},'' \href{http://dx.doi.org/10.1103/PhysRevD.80.035013}{{\em Phys.
  Rev.} {\bf D80} (2009)  035013},
\href{http://arxiv.org/abs/0905.3749}{{\tt arXiv:0905.3749 [hep-ph]}}.

\bibitem{Sannino:2009za}
F.~Sannino, ``{Conformal Dynamics for TeV Physics and Cosmology},'' {\em Acta
  Phys. Polon.} {\bf B40} (2009)  3533--3743,
\href{http://arxiv.org/abs/0911.0931}{{\tt arXiv:0911.0931 [hep-ph]}}.

\bibitem{Barbieri:2010mn}
R.~Barbieri, S.~Rychkov, and R.~Torre, ``{Signals of composite
  electroweak-neutral Dark Matter: LHC/Direct Detection interplay},''
  \href{http://dx.doi.org/10.1016/j.physletb.2010.04.010}{{\em Phys. Lett.}
  {\bf B688} (2010)  212--215},
\href{http://arxiv.org/abs/1001.3149}{{\tt arXiv:1001.3149 [hep-ph]}}.

\bibitem{Belyaev:2010kp}
A.~Belyaev, M.~T. Frandsen, S.~Sarkar, and F.~Sannino, ``{Mixed dark matter
  from technicolor},'' \href{http://dx.doi.org/10.1103/PhysRevD.83.015007}{{\em
  Phys. Rev.} {\bf D83} (2011)  015007},
\href{http://arxiv.org/abs/1007.4839}{{\tt arXiv:1007.4839 [hep-ph]}}.

\bibitem{Appelquist:2013ms}
{\bf Lattice Strong Dynamics (LSD) Collaboration} Collaboration, T.~Appelquist
  {\em et al.}, ``{Lattice calculation of composite dark matter form
  factors},'' \href{http://dx.doi.org/10.1103/PhysRevD.88.014502}{{\em
  Phys.Rev.} {\bf D88} (2013) no.~1, 014502},
\href{http://arxiv.org/abs/1301.1693}{{\tt arXiv:1301.1693 [hep-ph]}}.

\bibitem{Hietanen:2013fya}
A.~Hietanen, R.~Lewis, C.~Pica, and F.~Sannino, ``{Composite Goldstone Dark
  Matter: Experimental Predictions from the Lattice},''
\href{http://arxiv.org/abs/1308.4130}{{\tt arXiv:1308.4130 [hep-ph]}}.

\bibitem{Cline:2013zca}
J.~M. Cline, Z.~Liu, G.~Moore, and W.~Xue, ``{Composite strongly interacting
  dark matter},'' \href{http://dx.doi.org/10.1103/PhysRevD.90.015023}{{\em
  Phys. Rev.} {\bf D90} (2014) no.~1, 015023},
\href{http://arxiv.org/abs/1312.3325}{{\tt arXiv:1312.3325 [hep-ph]}}.

\bibitem{Appelquist:2014jch}
{\bf Lattice Strong Dynamics (LSD) Collaboration} Collaboration, T.~Appelquist
  {\em et al.}, ``{Composite bosonic baryon dark matter on the lattice: SU(4)
  baryon spectrum and the effective Higgs interaction},''
  \href{http://dx.doi.org/10.1103/PhysRevD.89.094508}{{\em Phys.Rev.} {\bf D89}
  (2014) no.~9, 094508},
\href{http://arxiv.org/abs/1402.6656}{{\tt arXiv:1402.6656 [hep-lat]}}.

\bibitem{Krnjaic:2014xza}
G.~Krnjaic and K.~Sigurdson, ``{Big Bang Darkleosynthesis},''
  \href{http://dx.doi.org/10.1016/j.physletb.2015.11.001}{{\em Phys. Lett.}
  {\bf B751} (2015)  464--468},
\href{http://arxiv.org/abs/1406.1171}{{\tt arXiv:1406.1171 [hep-ph]}}.

\bibitem{Detmold:2014qqa}
W.~Detmold, M.~McCullough, and A.~Pochinsky, ``{Dark Nuclei I: Cosmology and
  Indirect Detection},''
  \href{http://dx.doi.org/10.1103/PhysRevD.90.115013}{{\em Phys.Rev.} {\bf D90}
  (2014) no.~11, 115013},
\href{http://arxiv.org/abs/1406.2276}{{\tt arXiv:1406.2276 [hep-ph]}}.

\bibitem{Detmold:2014kba}
W.~Detmold, M.~McCullough, and A.~Pochinsky, ``{Dark nuclei. II. Nuclear
  spectroscopy in two-color QCD},''
  \href{http://dx.doi.org/10.1103/PhysRevD.90.114506}{{\em Phys.Rev.} {\bf D90}
  (2014) no.~11, 114506},
\href{http://arxiv.org/abs/1406.4116}{{\tt arXiv:1406.4116 [hep-lat]}}.

\bibitem{Brod:2014loa}
J.~Brod, J.~Drobnak, A.~L. Kagan, E.~Stamou, and J.~Zupan, ``{Stealth QCD-like
  strong interactions and the $t \bar {t}$ asymmetry},''
  \href{http://dx.doi.org/10.1103/PhysRevD.91.095009}{{\em Phys. Rev.} {\bf
  D91} (2015) no.~9, 095009},
\href{http://arxiv.org/abs/1407.8188}{{\tt arXiv:1407.8188 [hep-ph]}}.

\bibitem{Asano:2014wra}
M.~Asano and R.~Kitano, ``{Partially Composite Dark Matter},''
  \href{http://dx.doi.org/10.1007/JHEP09(2014)171}{{\em JHEP} {\bf 09} (2014)
  171},
\href{http://arxiv.org/abs/1406.6374}{{\tt arXiv:1406.6374 [hep-ph]}}.

\bibitem{Appelquist:2015yfa}
T.~Appelquist {\em et al.}, ``{Stealth Dark Matter: Dark scalar baryons through
  the Higgs portal},'' \href{http://dx.doi.org/10.1103/PhysRevD.92.075030}{{\em
  Phys. Rev.} {\bf D92} (2015) no.~7, 075030},
\href{http://arxiv.org/abs/1503.04203}{{\tt arXiv:1503.04203 [hep-ph]}}.

\bibitem{Appelquist:2015zfa}
T.~Appelquist {\em et al.}, ``{Detecting Stealth Dark Matter Directly through
  Electromagnetic Polarizability},''
  \href{http://dx.doi.org/10.1103/PhysRevLett.115.171803}{{\em Phys. Rev.
  Lett.} {\bf 115} (2015) no.~17, 171803},
\href{http://arxiv.org/abs/1503.04205}{{\tt arXiv:1503.04205 [hep-ph]}}.

\bibitem{Drach:2015epq}
V.~Drach, A.~Hietanen, C.~Pica, J.~Rantaharju, and F.~Sannino, ``{Template
  Composite Dark Matter: $SU(2)$ gauge theory with 2 fundamental flavours},''
  \href{http://dx.doi.org/10.22323/1.251.0234}{{\em PoS} {\bf LATTICE2015}
  (2016)  234},
\href{http://arxiv.org/abs/1511.04370}{{\tt arXiv:1511.04370 [hep-lat]}}.

\bibitem{Fichet:2016clq}
S.~Fichet, ``{Shining Light on Polarizable Dark Particles},''
  \href{http://dx.doi.org/10.1007/JHEP04(2017)088}{{\em JHEP} {\bf 04} (2017)
  088},
\href{http://arxiv.org/abs/1609.01762}{{\tt arXiv:1609.01762 [hep-ph]}}.

\bibitem{Co:2016akw}
R.~T. Co, K.~Harigaya, and Y.~Nomura, ``{Chiral Dark Sector},''
  \href{http://dx.doi.org/10.1103/PhysRevLett.118.101801}{{\em Phys. Rev.
  Lett.} {\bf 118} (2017) no.~10, 101801},
\href{http://arxiv.org/abs/1610.03848}{{\tt arXiv:1610.03848 [hep-ph]}}.

\bibitem{Dienes:2016vei}
K.~R. Dienes, F.~Huang, S.~Su, and B.~Thomas, ``{Dynamical Dark Matter from
  Strongly-Coupled Dark Sectors},''
  \href{http://dx.doi.org/10.1103/PhysRevD.95.043526}{{\em Phys. Rev.} {\bf
  D95} (2017) no.~4, 043526},
\href{http://arxiv.org/abs/1610.04112}{{\tt arXiv:1610.04112 [hep-ph]}}.

\bibitem{Ishida:2016fbp}
H.~Ishida, S.~Matsuzaki, and Y.~Yamaguchi, ``{Bosonic-Seesaw Portal Dark
  Matter},'' \href{http://dx.doi.org/10.1093/ptep/ptx132}{{\em PTEP} {\bf 2017}
  (2017) no.~10, 103B01},
\href{http://arxiv.org/abs/1610.07137}{{\tt arXiv:1610.07137 [hep-ph]}}.

\bibitem{Francis:2016bzf}
A.~Francis, R.~J. Hudspith, R.~Lewis, and S.~Tulin, ``{Dark matter from
  one-flavor SU(2) gauge theory},''
  \href{http://dx.doi.org/10.22323/1.256.0227}{{\em PoS} {\bf LATTICE2016}
  (2016)  227},
\href{http://arxiv.org/abs/1610.10068}{{\tt arXiv:1610.10068 [hep-lat]}}.

\bibitem{Lonsdale:2017mzg}
S.~J. Lonsdale, M.~Schroor, and R.~R. Volkas, ``{Asymmetric Dark Matter and the
  hadronic spectra of hidden QCD},''
  \href{http://dx.doi.org/10.1103/PhysRevD.96.055027}{{\em Phys. Rev.} {\bf
  D96} (2017) no.~5, 055027},
\href{http://arxiv.org/abs/1704.05213}{{\tt arXiv:1704.05213 [hep-ph]}}.

\bibitem{Berryman:2017twh}
J.~M. Berryman, A.~de~Gouvêa, K.~J. Kelly, and Y.~Zhang, ``{Dark Matter and
  Neutrino Mass from the Smallest Non-Abelian Chiral Dark Sector},''
  \href{http://dx.doi.org/10.1103/PhysRevD.96.075010}{{\em Phys. Rev.} {\bf
  D96} (2017) no.~7, 075010},
\href{http://arxiv.org/abs/1706.02722}{{\tt arXiv:1706.02722 [hep-ph]}}.

\bibitem{Mitridate:2017oky}
A.~Mitridate, M.~Redi, J.~Smirnov, and A.~Strumia, ``{Dark Matter as a weakly
  coupled Dark Baryon},'' \href{http://dx.doi.org/10.1007/JHEP10(2017)210}{{\em
  JHEP} {\bf 10} (2017)  210},
\href{http://arxiv.org/abs/1707.05380}{{\tt arXiv:1707.05380 [hep-ph]}}.

\bibitem{Francis:2018xjd}
A.~Francis, R.~J. Hudspith, R.~Lewis, and S.~Tulin, ``{Dark Matter from Strong
  Dynamics: The Minimal Theory of Dark Baryons},''
\href{http://arxiv.org/abs/1809.09117}{{\tt arXiv:1809.09117 [hep-ph]}}.

\bibitem{Kribs:2016cew}
G.~D. Kribs and E.~T. Neil, ``{Review of strongly-coupled composite dark matter
  models and lattice simulations},''
  \href{http://dx.doi.org/10.1142/S0217751X16430041}{{\em Int. J. Mod. Phys.}
  {\bf A31} (2016) no.~22, 1643004},
\href{http://arxiv.org/abs/1604.04627}{{\tt arXiv:1604.04627 [hep-ph]}}.

\bibitem{Kilic:2009mi}
C.~Kilic, T.~Okui, and R.~Sundrum, ``{Vectorlike Confinement at the LHC},''
  \href{http://dx.doi.org/10.1007/JHEP02(2010)018}{{\em JHEP} {\bf 02} (2010)
  018},
\href{http://arxiv.org/abs/0906.0577}{{\tt arXiv:0906.0577 [hep-ph]}}.

\bibitem{Kilic:2010et}
C.~Kilic and T.~Okui, ``{The LHC Phenomenology of Vectorlike Confinement},''
  \href{http://dx.doi.org/10.1007/JHEP04(2010)128}{{\em JHEP} {\bf 04} (2010)
  128},
\href{http://arxiv.org/abs/1001.4526}{{\tt arXiv:1001.4526 [hep-ph]}}.

\bibitem{Harnik:2011mv}
R.~Harnik, G.~D. Kribs, and A.~Martin, ``{Quirks at the Tevatron and Beyond},''
  \href{http://dx.doi.org/10.1103/PhysRevD.84.035029}{{\em Phys. Rev.} {\bf
  D84} (2011)  035029},
\href{http://arxiv.org/abs/1106.2569}{{\tt arXiv:1106.2569 [hep-ph]}}.

\bibitem{Fok:2011yc}
R.~Fok and G.~D. Kribs, ``{Chiral Quirkonium Decays},''
  \href{http://dx.doi.org/10.1103/PhysRevD.84.035001}{{\em Phys. Rev.} {\bf
  D84} (2011)  035001},
\href{http://arxiv.org/abs/1106.3101}{{\tt arXiv:1106.3101 [hep-ph]}}.

\bibitem{Bai:2013xga}
Y.~Bai and P.~Schwaller, ``{Scale of dark QCD},''
  \href{http://dx.doi.org/10.1103/PhysRevD.89.063522}{{\em Phys. Rev.} {\bf
  D89} (2014) no.~6, 063522},
\href{http://arxiv.org/abs/1306.4676}{{\tt arXiv:1306.4676 [hep-ph]}}.

\bibitem{Chacko:2015fbc}
Z.~Chacko, D.~Curtin, and C.~B. Verhaaren, ``{A Quirky Probe of Neutral
  Naturalness},'' \href{http://dx.doi.org/10.1103/PhysRevD.94.011504}{{\em
  Phys. Rev.} {\bf D94} (2016) no.~1, 011504},
\href{http://arxiv.org/abs/1512.05782}{{\tt arXiv:1512.05782 [hep-ph]}}.

\bibitem{Agashe:2016rle}
K.~Agashe, P.~Du, S.~Hong, and R.~Sundrum, ``{Flavor Universal Resonances and
  Warped Gravity},'' \href{http://dx.doi.org/10.1007/JHEP01(2017)016}{{\em
  JHEP} {\bf 01} (2017)  016},
\href{http://arxiv.org/abs/1608.00526}{{\tt arXiv:1608.00526 [hep-ph]}}.

\bibitem{Matsuzaki:2017bpp}
S.~Matsuzaki, K.~Nishiwaki, and R.~Watanabe, ``{Phenomenology of flavorful
  composite vector bosons in light of $B$ anomalies},''
  \href{http://dx.doi.org/10.1007/JHEP08(2017)145}{{\em JHEP} {\bf 08} (2017)
  145},
\href{http://arxiv.org/abs/1706.01463}{{\tt arXiv:1706.01463 [hep-ph]}}.

\bibitem{Draper:2018tmh}
P.~Draper, J.~Kozaczuk, and J.-H. Yu, ``{Theta in new QCD-like sectors},''
  \href{http://dx.doi.org/10.1103/PhysRevD.98.015028}{{\em Phys. Rev.} {\bf
  D98} (2018) no.~1, 015028},
\href{http://arxiv.org/abs/1803.00015}{{\tt arXiv:1803.00015 [hep-ph]}}.

\bibitem{Buttazzo:2018qqp}
D.~Buttazzo, D.~Redigolo, F.~Sala, and A.~Tesi, ``{Fusing Vectors into Scalars
  at High Energy Lepton Colliders},''
\href{http://arxiv.org/abs/1807.04743}{{\tt arXiv:1807.04743 [hep-ph]}}.

\bibitem{Schwaller:2015gea}
P.~Schwaller, D.~Stolarski, and A.~Weiler, ``{Emerging Jets},''
  \href{http://dx.doi.org/10.1007/JHEP05(2015)059}{{\em JHEP} {\bf 05} (2015)
  059},
\href{http://arxiv.org/abs/1502.05409}{{\tt arXiv:1502.05409 [hep-ph]}}.

\bibitem{Cohen:2015toa}
T.~Cohen, M.~Lisanti, and H.~K. Lou, ``{Semivisible Jets: Dark Matter
  Undercover at the LHC},''
  \href{http://dx.doi.org/10.1103/PhysRevLett.115.171804}{{\em Phys. Rev.
  Lett.} {\bf 115} (2015) no.~17, 171804},
\href{http://arxiv.org/abs/1503.00009}{{\tt arXiv:1503.00009 [hep-ph]}}.

\bibitem{Freytsis:2016dgf}
M.~Freytsis, S.~Knapen, D.~J. Robinson, and Y.~Tsai, ``{Gamma-rays from Dark
  Showers with Twin Higgs Models},''
  \href{http://dx.doi.org/10.1007/JHEP05(2016)018}{{\em JHEP} {\bf 05} (2016)
  018},
\href{http://arxiv.org/abs/1601.07556}{{\tt arXiv:1601.07556 [hep-ph]}}.

\bibitem{Zhang:2016sll}
M.~Kim, H.-S. Lee, M.~Park, and M.~Zhang, ``{Examining the origin of dark
  matter mass at colliders},''
\href{http://arxiv.org/abs/1612.02850}{{\tt arXiv:1612.02850 [hep-ph]}}.

\bibitem{Cohen:2017pzm}
T.~Cohen, M.~Lisanti, H.~K. Lou, and S.~Mishra-Sharma, ``{LHC Searches for Dark
  Sector Showers},'' \href{http://dx.doi.org/10.1007/JHEP11(2017)196}{{\em
  JHEP} {\bf 11} (2017)  196},
\href{http://arxiv.org/abs/1707.05326}{{\tt arXiv:1707.05326 [hep-ph]}}.

\bibitem{Beauchesne:2017yhh}
H.~Beauchesne, E.~Bertuzzo, G.~Grilli Di~Cortona, and Z.~Tabrizi, ``{Collider
  phenomenology of Hidden Valley mediators of spin 0 or 1/2 with semivisible
  jets},'' \href{http://dx.doi.org/10.1007/JHEP08(2018)030}{{\em JHEP} {\bf 08}
  (2018)  030},
\href{http://arxiv.org/abs/1712.07160}{{\tt arXiv:1712.07160 [hep-ph]}}.

\bibitem{Renner:2018fhh}
S.~Renner and P.~Schwaller, ``{A flavoured dark sector},''
  \href{http://dx.doi.org/10.1007/JHEP08(2018)052}{{\em JHEP} {\bf 08} (2018)
  052},
\href{http://arxiv.org/abs/1803.08080}{{\tt arXiv:1803.08080 [hep-ph]}}.

\bibitem{Mahbubani:2017gjh}
R.~Mahbubani, P.~Schwaller, and J.~Zurita, ``{Closing the window for compressed
  Dark Sectors with disappearing charged tracks},''
  \href{http://dx.doi.org/10.1007/JHEP06(2017)119,
  10.1007/JHEP10(2017)061}{{\em JHEP} {\bf 06} (2017)  119},
  \href{http://arxiv.org/abs/1703.05327}{{\tt arXiv:1703.05327 [hep-ph]}}.
[Erratum: JHEP10,061(2017)].

\bibitem{Buchmueller:2017uqu}
O.~Buchmueller, A.~De~Roeck, K.~Hahn, M.~McCullough, P.~Schwaller, K.~Sung, and
  T.-T. Yu, ``{Simplified Models for Displaced Dark Matter Signatures},''
  \href{http://dx.doi.org/10.1007/JHEP09(2017)076}{{\em JHEP} {\bf 09} (2017)
  076},
\href{http://arxiv.org/abs/1704.06515}{{\tt arXiv:1704.06515 [hep-ph]}}.

\bibitem{Daci:2015hca}
N.~Daci, I.~De~Bruyn, S.~Lowette, M.~H.~G. Tytgat, and B.~Zaldivar,
  ``{Simplified SIMPs and the LHC},''
  \href{http://dx.doi.org/10.1007/JHEP11(2015)108}{{\em JHEP} {\bf 11} (2015)
  108},
\href{http://arxiv.org/abs/1503.05505}{{\tt arXiv:1503.05505 [hep-ph]}}.

\bibitem{Hochberg:2017khi}
Y.~Hochberg, E.~Kuflik, and H.~Murayama, ``{Dark spectroscopy at lepton
  colliders},'' \href{http://dx.doi.org/10.1103/PhysRevD.97.055030}{{\em Phys.
  Rev.} {\bf D97} (2018) no.~5, 055030},
\href{http://arxiv.org/abs/1706.05008}{{\tt arXiv:1706.05008 [hep-ph]}}.

\bibitem{Han:2007ae}
T.~Han, Z.~Si, K.~M. Zurek, and M.~J. Strassler, ``{Phenomenology of hidden
  valleys at hadron colliders},''
  \href{http://dx.doi.org/10.1088/1126-6708/2008/07/008}{{\em JHEP} {\bf 07}
  (2008)  008},
\href{http://arxiv.org/abs/0712.2041}{{\tt arXiv:0712.2041 [hep-ph]}}.

\bibitem{Kang:2008ea}
J.~Kang and M.~A. Luty, ``{Macroscopic Strings and 'Quirks' at Colliders},''
  \href{http://dx.doi.org/10.1088/1126-6708/2009/11/065}{{\em JHEP} {\bf 11}
  (2009)  065},
\href{http://arxiv.org/abs/0805.4642}{{\tt arXiv:0805.4642 [hep-ph]}}.

\bibitem{Harnik:2008ax}
R.~Harnik and T.~Wizansky, ``{Signals of New Physics in the Underlying
  Event},'' \href{http://dx.doi.org/10.1103/PhysRevD.80.075015}{{\em Phys.
  Rev.} {\bf D80} (2009)  075015},
\href{http://arxiv.org/abs/0810.3948}{{\tt arXiv:0810.3948 [hep-ph]}}.

\bibitem{Knapen:2016hky}
S.~Knapen, S.~Pagan~Griso, M.~Papucci, and D.~J. Robinson, ``{Triggering Soft
  Bombs at the LHC},'' \href{http://dx.doi.org/10.1007/JHEP08(2017)076}{{\em
  JHEP} {\bf 08} (2017)  076},
\href{http://arxiv.org/abs/1612.00850}{{\tt arXiv:1612.00850 [hep-ph]}}.

\bibitem{Pierce:2017taw}
A.~Pierce, B.~Shakya, Y.~Tsai, and Y.~Zhao, ``{Searching for confining hidden
  valleys at LHCb, ATLAS, and CMS},''
  \href{http://dx.doi.org/10.1103/PhysRevD.97.095033}{{\em Phys. Rev.} {\bf
  D97} (2018) no.~9, 095033},
\href{http://arxiv.org/abs/1708.05389}{{\tt arXiv:1708.05389 [hep-ph]}}.

\bibitem{Kribs:2018oad}
G.~D. Kribs, A.~Martin, and T.~Tong, ``{Effective Theories of Dark Mesons with
  Custodial Symmetry},''
\href{http://arxiv.org/abs/1809.10183}{{\tt arXiv:1809.10183 [hep-ph]}}.

\bibitem{Beylin:2016kga}
V.~Beylin, M.~Bezuglov, V.~Kuksa, and N.~Volchanskiy, ``{An analysis of a
  minimal vectorlike extension of the Standard Model},''
  \href{http://dx.doi.org/10.1155/2017/1765340}{{\em Adv. High Energy Phys.}
  {\bf 2017} (2017)  1765340},
\href{http://arxiv.org/abs/1611.06006}{{\tt arXiv:1611.06006 [hep-ph]}}.

\bibitem{Ecker:1988te}
G.~Ecker, J.~Gasser, A.~Pich, and E.~de~Rafael, ``{The Role of Resonances in
  Chiral Perturbation Theory},''
\href{http://dx.doi.org/10.1016/0550-3213(89)90346-5}{{\em Nucl. Phys.} {\bf
  B321} (1989)  311--342}.

\bibitem{Ecker:1989yg}
G.~Ecker, J.~Gasser, H.~Leutwyler, A.~Pich, and E.~de~Rafael, ``{Chiral
  Lagrangians for Massive Spin 1 Fields},''
\href{http://dx.doi.org/10.1016/0370-2693(89)91627-4}{{\em Phys. Lett.} {\bf
  B223} (1989)  425--432}.

\bibitem{Alexander:2016aln}
J.~Alexander {\em et al.}, ``{Dark Sectors 2016 Workshop: Community Report},''
\newblock 2016.
\newblock \href{http://arxiv.org/abs/1608.08632}{{\tt arXiv:1608.08632
  [hep-ph]}}.
\newblock
\url{http://lss.fnal.gov/archive/2016/conf/fermilab-conf-16-421.pdf}.
\newblock

\bibitem{Lee:2018pag}
L.~Lee, C.~Ohm, A.~Soffer, and T.-T. Yu, ``{Collider Searches for Long-Lived
  Particles Beyond the Standard Model},''
\href{http://arxiv.org/abs/1810.12602}{{\tt arXiv:1810.12602 [hep-ph]}}.

\bibitem{Gunion:1989we}
J.~F. Gunion, H.~E. Haber, G.~L. Kane, and S.~Dawson, ``{The Higgs Hunter's
  Guide},''
{\em Front. Phys.} {\bf 80} (2000)  1--404.

\bibitem{Alloul:2013bka}
A.~Alloul, N.~D. Christensen, C.~Degrande, C.~Duhr, and B.~Fuks, ``{FeynRules
  2.0 - A complete toolbox for tree-level phenomenology},''
  \href{http://dx.doi.org/10.1016/j.cpc.2014.04.012}{{\em Comput. Phys.
  Commun.} {\bf 185} (2014)  2250--2300},
\href{http://arxiv.org/abs/1310.1921}{{\tt arXiv:1310.1921 [hep-ph]}}.

\bibitem{Alwall:2014hca}
J.~Alwall, R.~Frederix, S.~Frixione, V.~Hirschi, F.~Maltoni, O.~Mattelaer,
  H.~S. Shao, T.~Stelzer, P.~Torrielli, and M.~Zaro, ``{The automated
  computation of tree-level and next-to-leading order differential cross
  sections, and their matching to parton shower simulations},''
  \href{http://dx.doi.org/10.1007/JHEP07(2014)079}{{\em JHEP} {\bf 07} (2014)
  079},
\href{http://arxiv.org/abs/1405.0301}{{\tt arXiv:1405.0301 [hep-ph]}}.

\bibitem{Sjostrand:2014zea}
T.~Sjöstrand, S.~Ask, J.~R. Christiansen, R.~Corke, N.~Desai, P.~Ilten,
  S.~Mrenna, S.~Prestel, C.~O. Rasmussen, and P.~Z. Skands, ``{An Introduction
  to PYTHIA 8.2},'' \href{http://dx.doi.org/10.1016/j.cpc.2015.01.024}{{\em
  Comput. Phys. Commun.} {\bf 191} (2015)  159--177},
\href{http://arxiv.org/abs/1410.3012}{{\tt arXiv:1410.3012 [hep-ph]}}.

\bibitem{deFavereau:2013fsa}
{\bf DELPHES 3} Collaboration, J.~de~Favereau, C.~Delaere, P.~Demin,
  A.~Giammanco, V.~Lemaître, A.~Mertens, and M.~Selvaggi, ``{DELPHES 3, A
  modular framework for fast simulation of a generic collider experiment},''
  \href{http://dx.doi.org/10.1007/JHEP02(2014)057}{{\em JHEP} {\bf 02} (2014)
  057},
\href{http://arxiv.org/abs/1307.6346}{{\tt arXiv:1307.6346 [hep-ex]}}.

\bibitem{Cacciari:2011ma}
M.~Cacciari, G.~P. Salam, and G.~Soyez, ``{FastJet User Manual},''
  \href{http://dx.doi.org/10.1140/epjc/s10052-012-1896-2}{{\em Eur. Phys. J.}
  {\bf C72} (2012)  1896},
\href{http://arxiv.org/abs/1111.6097}{{\tt arXiv:1111.6097 [hep-ph]}}.

\bibitem{Cacciari:2008gp}
M.~Cacciari, G.~P. Salam, and G.~Soyez, ``{The Anti-k(t) jet clustering
  algorithm},'' \href{http://dx.doi.org/10.1088/1126-6708/2008/04/063}{{\em
  JHEP} {\bf 04} (2008)  063},
\href{http://arxiv.org/abs/0802.1189}{{\tt arXiv:0802.1189 [hep-ph]}}.

\bibitem{Aaboud:2017buh}
{\bf ATLAS} Collaboration, M.~Aaboud {\em et al.}, ``{Search for new high-mass
  phenomena in the dilepton final state using 36 fb$^{-1}$ of proton-proton
  collision data at $\sqrt{s}=13 $ TeV with the ATLAS detector},''
  \href{http://dx.doi.org/10.1007/JHEP10(2017)182}{{\em JHEP} {\bf 10} (2017)
  182},
\href{http://arxiv.org/abs/1707.02424}{{\tt arXiv:1707.02424 [hep-ex]}}.

\bibitem{Sirunyan:2018exx}
{\bf CMS} Collaboration, A.~M. Sirunyan {\em et al.}, ``{Search for high-mass
  resonances in dilepton final states in proton-proton collisions at
  $\sqrt{s}=$ 13 TeV},''
\href{http://arxiv.org/abs/1803.06292}{{\tt arXiv:1803.06292 [hep-ex]}}.

\bibitem{HEPDATA_DILEPTON}
{\bf ATLAS} Collaboration, ``{Search for new high-mass phenomena in the
  dilepton final state using 36.1 fb$^{-1}$ of proton-proton collision data at
  $\sqrt{s}$ = 13 TeV with the ATLAS detector},'' 2017.
\newblock
  \url{http://www.hepdata.net/record/ins1609250?version=1&table=Table6}.

\bibitem{Schael:2013ita}
{\bf DELPHI, OPAL, LEP Electroweak, ALEPH, L3} Collaboration, S.~Schael {\em et
  al.}, ``{Electroweak Measurements in Electron-Positron Collisions at
  W-Boson-Pair Energies at LEP},''
  \href{http://dx.doi.org/10.1016/j.physrep.2013.07.004}{{\em Phys. Rept.} {\bf
  532} (2013)  119--244},
\href{http://arxiv.org/abs/1302.3415}{{\tt arXiv:1302.3415 [hep-ex]}}.

\bibitem{Aaltonen:2014hua}
{\bf CDF} Collaboration, T.~A. Aaltonen {\em et al.}, ``{Study of Top-Quark
  Production and Decays involving a Tau Lepton at CDF and Limits on a
  Charged-Higgs Boson Contribution},''
  \href{http://dx.doi.org/10.1103/PhysRevD.89.091101}{{\em Phys. Rev.} {\bf
  D89} (2014) no.~9, 091101},
\href{http://arxiv.org/abs/1402.6728}{{\tt arXiv:1402.6728 [hep-ex]}}.

\bibitem{Aad:2015dya}
{\bf ATLAS} Collaboration, G.~Aad {\em et al.}, ``{Measurements of the top
  quark branching ratios into channels with leptons and quarks with the ATLAS
  detector},'' \href{http://dx.doi.org/10.1103/PhysRevD.92.072005}{{\em Phys.
  Rev.} {\bf D92} (2015) no.~7, 072005},
\href{http://arxiv.org/abs/1506.05074}{{\tt arXiv:1506.05074 [hep-ex]}}.

\bibitem{Tanabashi:2018oca}
{\bf Particle Data Group} Collaboration, M.~Tanabashi {\em et al.}, ``{Review
  of Particle Physics},''
\href{http://dx.doi.org/10.1103/PhysRevD.98.030001}{{\em Phys. Rev.} {\bf D98}
  (2018) no.~3, 030001}.

\bibitem{Abazov:2012vd}
{\bf D0} Collaboration, V.~M. Abazov {\em et al.}, ``{An Improved determination
  of the width of the top quark},''
  \href{http://dx.doi.org/10.1103/PhysRevD.85.091104}{{\em Phys. Rev.} {\bf
  D85} (2012)  091104},
\href{http://arxiv.org/abs/1201.4156}{{\tt arXiv:1201.4156 [hep-ex]}}.

\bibitem{Khachatryan:2014nda}
{\bf CMS} Collaboration, V.~Khachatryan {\em et al.}, ``{Measurement of the
  ratio $\mathcal B(t \to Wb)/\mathcal B(t \to Wq)$ in pp collisions at
  $\sqrt{s}$ = 8 TeV},''
  \href{http://dx.doi.org/10.1016/j.physletb.2014.06.076}{{\em Phys. Lett.}
  {\bf B736} (2014)  33--57},
\href{http://arxiv.org/abs/1404.2292}{{\tt arXiv:1404.2292 [hep-ex]}}.

\bibitem{ATLAS:2016grc}
{\bf ATLAS} Collaboration, T.~A. collaboration,
``{Search for charged Higgs bosons in the $\tau$+jets final state using 14.7
  fb$^{-1}$ of pp collision data recorded at $\sqrt{s}=13$ TeV with the ATLAS
  experiment},''.

\bibitem{ATLAS:2016qiq}
{\bf ATLAS} Collaboration, T.~A. collaboration,
``{Search for charged Higgs bosons in the $H^{\pm}\to tb$ decay channel in $pp$
  collisions at $\sqrt{s}=13$ TeV using the ATLAS detector},''.

\bibitem{Aaboud:2017sjh}
{\bf ATLAS} Collaboration, M.~Aaboud {\em et al.}, ``{Search for additional
  heavy neutral Higgs and gauge bosons in the ditau final state produced in 36
  fb$^{-1}$ of pp collisions at $ \sqrt{s}=13 $ TeV with the ATLAS detector},''
  \href{http://dx.doi.org/10.1007/JHEP01(2018)055}{{\em JHEP} {\bf 01} (2018)
  055},
\href{http://arxiv.org/abs/1709.07242}{{\tt arXiv:1709.07242 [hep-ex]}}.

\bibitem{Liu:2015bma}
Z.~Liu and B.~Tweedie, ``{The Fate of Long-Lived Superparticles with Hadronic
  Decays after LHC Run 1},''
  \href{http://dx.doi.org/10.1007/JHEP06(2015)042}{{\em JHEP} {\bf 06} (2015)
  042},
\href{http://arxiv.org/abs/1503.05923}{{\tt arXiv:1503.05923 [hep-ph]}}.

\bibitem{Evans:2016zau}
J.~A. Evans and J.~Shelton, ``{Long-Lived Staus and Displaced Leptons at the
  LHC},'' \href{http://dx.doi.org/10.1007/JHEP04(2016)056}{{\em JHEP} {\bf 04}
  (2016)  056},
\href{http://arxiv.org/abs/1601.01326}{{\tt arXiv:1601.01326 [hep-ph]}}.

\bibitem{Curtin:2018mvb}
D.~Curtin {\em et al.}, ``{Long-Lived Particles at the Energy Frontier: The
  MATHUSLA Physics Case},''
\href{http://arxiv.org/abs/1806.07396}{{\tt arXiv:1806.07396 [hep-ph]}}.

\bibitem{Liu:2018wte}
J.~Liu, Z.~Liu, and L.-T. Wang, ``{Long-lived particles at the LHC: catching
  them in time},''
\href{http://arxiv.org/abs/1805.05957}{{\tt arXiv:1805.05957 [hep-ph]}}.

\bibitem{Barbieri:2007bh}
R.~Barbieri, B.~Bellazzini, V.~S. Rychkov, and A.~Varagnolo, ``{The Higgs boson
  from an extended symmetry},''
  \href{http://dx.doi.org/10.1103/PhysRevD.76.115008}{{\em Phys. Rev.} {\bf
  D76} (2007)  115008},
\href{http://arxiv.org/abs/0706.0432}{{\tt arXiv:0706.0432 [hep-ph]}}.

\bibitem{ATLAS-CONF-2016-093}
{\bf ATLAS Collaboration} Collaboration, ``{Search for electroweak production
  of supersymmetric particles in final states with tau leptons in $\sqrt{s} =$
  13TeV pp collisions with the ATLAS detector},'' Tech. Rep.
  ATLAS-CONF-2016-093, CERN, Geneva, Aug, 2016.
\newblock \url{http://cds.cern.ch/record/2211437}.

\bibitem{Aad:2014hja}
{\bf ATLAS} Collaboration, G.~Aad {\em et al.}, ``{Search for new phenomena in
  events with three or more charged leptons in $pp$ collisions at $\sqrt{s}=8$
  TeV with the ATLAS detector},''
  \href{http://dx.doi.org/10.1007/JHEP08(2015)138}{{\em JHEP} {\bf 08} (2015)
  138},
\href{http://arxiv.org/abs/1411.2921}{{\tt arXiv:1411.2921 [hep-ex]}}.

\bibitem{ATLAS-CONF-2017-029}
{\bf ATLAS Collaboration} Collaboration, ``{Measurement of the tau lepton
  reconstruction and identification performance in the ATLAS experiment using
  $pp$ collisions at $\sqrt{s}=13~{\rm TeV}$},'' Tech. Rep.
  ATLAS-CONF-2017-029, CERN, Geneva, May, 2017.
\newblock \url{https://cds.cern.ch/record/2261772}.

\bibitem{Chatrchyan:2014aea}
{\bf CMS} Collaboration, S.~Chatrchyan {\em et al.}, ``{Search for anomalous
  production of events with three or more leptons in $pp$ collisions at
  $\sqrt(s) =$ 8 TeV},''
  \href{http://dx.doi.org/10.1103/PhysRevD.90.032006}{{\em Phys. Rev.} {\bf
  D90} (2014)  032006},
\href{http://arxiv.org/abs/1404.5801}{{\tt arXiv:1404.5801 [hep-ex]}}.

\bibitem{ATLAS:2013tma}
{\bf ATLAS} Collaboration, ``{Search for strongly produced superpartners in
  final states with two same sign leptons with the ATLAS detector using 21 fb-1
  of proton-proton collisions at sqrt(s)=8 TeV.},'' Tech. Rep.
  ATLAS-CONF-2013-007, 2013.

\bibitem{Aaboud:2017dmy}
{\bf ATLAS} Collaboration, M.~Aaboud {\em et al.}, ``{Search for supersymmetry
  in final states with two same-sign or three leptons and jets using 36
  fb$^{-1}$ of $\sqrt{s}=13$ TeV $pp$ collision data with the ATLAS
  detector},'' \href{http://dx.doi.org/10.1007/JHEP09(2017)084}{{\em JHEP} {\bf
  09} (2017)  084},
\href{http://arxiv.org/abs/1706.03731}{{\tt arXiv:1706.03731 [hep-ex]}}.

\bibitem{Aad:2015wra}
{\bf ATLAS} Collaboration, G.~Aad {\em et al.}, ``{Search for a CP-odd Higgs
  boson decaying to Zh in pp collisions at $\sqrt{s} = 8$ TeV with the ATLAS
  detector},'' \href{http://dx.doi.org/10.1016/j.physletb.2015.03.054}{{\em
  Phys. Lett.} {\bf B744} (2015)  163--183},
\href{http://arxiv.org/abs/1502.04478}{{\tt arXiv:1502.04478 [hep-ex]}}.

\bibitem{Aad:2015fna}
{\bf ATLAS} Collaboration, G.~Aad {\em et al.}, ``{A search for $ t\overline{t}
  $ resonances using lepton-plus-jets events in proton-proton collisions at $
  \sqrt{s}=8 $ TeV with the ATLAS detector},''
  \href{http://dx.doi.org/10.1007/JHEP08(2015)148}{{\em JHEP} {\bf 08} (2015)
  148},
\href{http://arxiv.org/abs/1505.07018}{{\tt arXiv:1505.07018 [hep-ex]}}.

\bibitem{Khachatryan:2014ura}
{\bf CMS} Collaboration, V.~Khachatryan {\em et al.}, ``{Search for pair
  production of third-generation scalar leptoquarks and top squarks in
  proton–proton collisions at $\sqrt{s}$=8 TeV},''
  \href{http://dx.doi.org/10.1016/j.physletb.2014.10.063}{{\em Phys. Lett.}
  {\bf B739} (2014)  229--249},
\href{http://arxiv.org/abs/1408.0806}{{\tt arXiv:1408.0806 [hep-ex]}}.

\bibitem{Aaboud:2017hrg}
{\bf ATLAS} Collaboration, M.~Aaboud {\em et al.}, ``{Search for Supersymmetry
  in final states with missing transverse momentum and multiple $b$-jets in
  proton--proton collisions at $\sqrt{s} = 13$ TeV with the ATLAS detector},''
\href{http://arxiv.org/abs/1711.01901}{{\tt arXiv:1711.01901 [hep-ex]}}.

\bibitem{Khachatryan:2016cfx}
{\bf CMS} Collaboration, V.~Khachatryan {\em et al.}, ``{Search for heavy
  resonances decaying into a vector boson and a Higgs boson in final states
  with charged leptons, neutrinos, and b quarks},'' {\em Submitted to: Phys.
  Lett. B} (2016)  ,
\href{http://arxiv.org/abs/1610.08066}{{\tt arXiv:1610.08066 [hep-ex]}}.

\bibitem{Sirunyan:2017djm}
{\bf CMS} Collaboration, A.~M. Sirunyan {\em et al.}, ``{Search for Higgs boson
  pair production in events with two bottom quarks and two tau leptons in
  proton–proton collisions at $\sqrt s$ =13TeV},''
  \href{http://dx.doi.org/10.1016/j.physletb.2018.01.001}{{\em Phys. Lett.}
  {\bf B778} (2018)  101--127},
\href{http://arxiv.org/abs/1707.02909}{{\tt arXiv:1707.02909 [hep-ex]}}.

\bibitem{Sirunyan:2017nvi}
{\bf CMS} Collaboration, A.~M. Sirunyan {\em et al.}, ``{Search for low mass
  vector resonances decaying into quark-antiquark pairs in proton-proton
  collisions at $ \sqrt{s}=13 $ TeV},''
  \href{http://dx.doi.org/10.1007/JHEP01(2018)097}{{\em JHEP} {\bf 01} (2018)
  097},
\href{http://arxiv.org/abs/1710.00159}{{\tt arXiv:1710.00159 [hep-ex]}}.

\bibitem{Khachatryan:2016vph}
{\bf CMS} Collaboration, V.~Khachatryan {\em et al.}, ``{Search for single
  production of a heavy vector-like T quark decaying to a Higgs boson and a top
  quark with a lepton and jets in the final state},''
  \href{http://dx.doi.org/10.1016/j.physletb.2017.05.019}{{\em Phys. Lett.}
  {\bf B771} (2017)  80--105},
\href{http://arxiv.org/abs/1612.00999}{{\tt arXiv:1612.00999 [hep-ex]}}.

\end{thebibliography}\endgroup

\end{document}